\documentclass[a4paper,12pt,english]{article}
\pdfoutput=1
\usepackage[bindingoffset=0.3cm,textheight=22.5cm,hdivide={2.7cm,*,2.7cm}, 
vdivide={*,22cm,*}]{geometry}
\usepackage{cite}
\usepackage{amsmath,amsfonts,amssymb,babel,slashed,graphicx,color,empheq}
\usepackage[matrix,arrow]{xy}
\input{xy}
\xyoption{all}
\usepackage{caption}
\usepackage{subcaption}

\usepackage[bookmarksnumbered=true,breaklinks=true]{hyperref}



\newcommand{\del}{\partial}
\newcommand{\End}{\mathrm{End}}

\def\nn{\nonumber} \def \bea{\begin{eqnarray}} \def\eea{\end{eqnarray}}

\numberwithin{equation}{section}

\def\a{\alpha}  \def\b{\beta}
 \def\g{\gamma} 
 \def\d{\delta} 
    \def\k{\kappa}
\def\l{\lambda} \def\L{\Lambda}  \def\m{\mu}
    \def\r{\rho}
\def\s{\sigma}  

\def\cA{{\cal A}}  \def\cC{{\cal C}} 
 \def\cE{{\cal E}}  
 \def\cH{{\cal H}}  
 \def\cK{{\cal K}}  
\def\cM{{\cal M}}  \def\cO{{\cal O}} 
  \def\cR{{\cal R}} 
\def\cS{{\cal S}} \def\cT{{\cal T}} 
 \def\cW{{\cal W}}  
\def\cY{{\cal Y}}

\def\R{{\mathbb R}} \def\C{{\mathbb C}} 

 \def\one{\mbox{1 \kern-.59em {\rm l}}}

\def\mh{\mathfrak{h}}

\def\msu{\mathfrak{su}}
\def\mso{\mathfrak{so}}

\def\m{{\bf m}}
\def\t{{\bf t}}

\def\Tr{\mbox{Tr}}

\def\({\left(} \def\){\right)} \def\tens{\otimes}

\def\diag{\mathrm{diag}}

 \allowdisplaybreaks[3]

\begin{document}

\makeatother


\parindent=0cm

\renewcommand{\title}[1]{\vspace{10mm}\noindent{\Large{\bf

#1}}\vspace{8mm}} \newcommand{\authors}[1]{\noindent{\large

#1}\vspace{5mm}} \newcommand{\address}[1]{{\itshape #1\vspace{2mm}}}


\begin{titlepage}
\begin{flushright}
 UWThPh-2017-08 
\end{flushright}
\begin{center}
\title{ {\Large Covariant 4-dimensional fuzzy spheres, \\[1ex] matrix models and higher spin}  }

\vskip 3mm

\authors{Marcus Sperling{\footnote{marcus.sperling@univie.ac.at}} and Harold C. Steinacker{\footnote{harold.steinacker@univie.ac.at}}}

\vskip 3mm

 \address{ 
{\it Faculty of Physics, University of Vienna\\
Boltzmanngasse 5, A-1090 Vienna, Austria  }  
  }

\bigskip

\vskip 1.4cm

\textbf{Abstract}
\vskip 3mm

\begin{minipage}{14cm}%

We study in detail generalized 4-dimensional fuzzy spheres with twisted extra dimensions.
These spheres can be viewed as $SO(5)$-equivariant projections of quantized coadjoint orbits of $SO(6)$.
We show that they arise as solutions in Yang-Mills matrix models, which naturally leads 
to higher-spin gauge theories on $S^4$. 
Several types of embeddings in matrix models are found, including one with self-intersecting 
fuzzy extra dimensions $S^4 \times \cK$, which is expected to entail 2+1 generations.

\end{minipage}

\end{center}

\end{titlepage}

\tableofcontents

\section{Introduction}

The purpose of this paper is to present in detail a new class of covariant 4-dimensional quantum spheres $\cS^4_\L$,
and to show that they arise as solutions of Yang-Mills-type matrix models. 
The results in this paper should provide the basis  
for studying physical applications, in particular for
the higher spin theories arising on such spaces,
which were argued to contain gravity \cite{Steinacker:2016vgf,Heckman:2014xha}. 

The long-term  motivation for this work is to find an appropriate 
framework for quantum geometry which
accommodates both gravity and quantum mechanics. Among the many attempts towards 
this goal, one strategy is to 
use some sort of non-commutative geometry, in order to avoid point-like 
singularities. 
However, several problems arise in that context.
One issue is to ensure (local) Lorentz invariance, which must hold with high 
precision. Another issue is known as 
UV/IR mixing, which is a very generic phenomenon on non-commutative spaces that 
typically leads to unacceptably large non-local 
(and Lorentz-violating) effects 
\cite{Minwalla:1999px,Iso:2000ew,Kinar:2001yk,Steinacker:2016nsc}.
The latter problem is avoided in the  maximally supersymmetric IKKT model, which was 
proposed as a constructive definition 
of IIB string theory \cite{Ishibashi:1996xs}.

To address the issue of Lorentz invariance, it seems useful to examine in more 
detail the few known examples of 4-dimensional quantum (or fuzzy) spaces 
which are compatible with the full classical rotational and translational 
symmetries.
In the Euclidean case, the best-known example is the fuzzy 4-sphere $\cS^4_N$ 
\cite{Grosse:1996mz,Castelino:1997rv,Ramgoolam:2001zx}. 
Taking this as a background in the IKKT model 
indeed leads to an interesting  $SO(5)$-covariant higher-spin gauge theory
\cite{Steinacker:2016vgf}, see also \cite{Heckman:2014xha}.
However some issues with the proper identification of the graviton suggest 
to consider certain generalized fuzzy 4-spheres, which were introduced in \cite{Steinacker:2016vgf} 
and denoted as $\cS^4_\L$. These provide additional structure which may be
relevant\footnote{The basic fuzzy 4-sphere $\cS^4_N$  has also been discussed in other contexts including 
string theory \cite{Castelino:1997rv,Ho:2000br,Constable:2001ag,Balasubramanian:2001nh}, 
non-commutative field theory 
\cite{Grosse:1996mz,Kimura:2002nq,Medina:2002pc,Medina:2012cs,Azuma:2004yg}
and the generalized quantum Hall effect \cite{Zhang:2001xs,Karabali:2002im}.} to particle physics and gravity.
However, the internal structure of $\cS^4_\L$ is rather intricate 
and requires a detailed elaboration, which is provided in the present paper.

The most interesting feature of the generalized $\cS^4_\L$ is that they realize 
a compactified phase space (or tangent bundle) of a 4-sphere, supplemented by 
additional structure. They are in fact 10-dimensional bundles over $S^4$, which 
allows to naturally accommodate graviton modes as coefficients of momentum 
generators, 
as discussed in \cite{Steinacker:2016vgf}. 
Additionally, the structure of $\cS^4_\L$ allows for novel types of 
embeddings in Yang-Mills matrix models, 
using either position generators, or momentum generators, or both.
In the latter case, the fuzzy extra dimensions provide further structure
and should justify the dimensional reduction to $S^4$.
Finally, the momentum picture, as suggested in \cite{Hanada:2005vr}, can now 
be realized in a well-defined way via finite-dimensional matrices,
which should allow to clarify the mechanism for gravity in this scenario.

The mathematical structure of the novel class of spaces is as follows: 
Fuzzy $\cS^4_\L$ is a quantization of a certain coadjoint orbit $\cO_\L$ of 
$SO(6)$, defined in terms of irreducible representations with highest weight 
$\L$.
The orbit $\cO_\L$ is a bundle over $S^4$ via some $SO(5)$-covariant 
projection, similar to a Hopf map. 
Since the underlying orbit  $\cO_\L$ is  10-dimensional (or 
6-dimensional for fuzzy $\cS^4_N$),
there is a 6-dimensional fiber, and $\cO_\L$ is a twisted (=equivariant) bundle over 
$S^4$. The ``twisted'' bundle structure means that the 
local stabilizer group $SO(4)$ of some point on $S^4$ acts non-trivially on the 
fiber. Consequently, fluctuations lead to a higher spin theory on $S^4$, in 
marked contrast to conventional Kaluza-Klein compactifications.

The accidental isomorphisms $\mso(6) \cong \msu(4)$ provides two useful and 
complementary views of $\cO_\L$:
either as a $S^4 \times S^2$ bundle over $S^4$ or as a $\C P^2$ bundle over $\C 
P^3$, which in turn is a $S^2$ bundle over $S^4$.
The first picture is compatible with $SO(5)$, but some degeneracies are hidden, 
which are resolved in the second picture.
In particular, we observe an interesting triple self-intersection (or covering) 
structure in the extra dimensions of $\cS^4_\L$, which is best 
understood in terms of the $\C P^2$ fiber. This is very similar 
to the squashed $\C P^2$ 
found as extra dimensions 
in \cite{Steinacker:2014lma,Steinacker:2015mia}, and it strongly suggests 3 
families of fermionic (near-)zero modes. 
In fact, one of these families would be distinct from the other two, 
which suggests a ``2+1'' family structure. This is a very intriguing albeit 
preliminary observation, 
and a further elaboration is postponed to future work.

The outline of this article is as follows: first, we briefly review the basic 
fuzzy 4-sphere in Section \ref{sec:fuzzy_4-sphere}. Next, in order to describe 
the aforementioned structures of the generalized 4-sphere explicitly, we 
identify suitable matrix coordinates in the semi-classical case in Section 
\ref{sec:twisted_bdl_coadj_orbit}.
The semi-classical Poisson geometry then suggests the appropriate generators for 
the fuzzy case,
and in Section \ref{sec:fuzzy_geometry} we work out the algebraic properties of 
these matrix observables. 
In particular, we show that the generalized $\cS^4_\L$ is indeed a solution of the 
Yang-Mills type matrix models such as the IKKT model, with a suitable potential 
term
that is argued to arise at 1-loop. Apart from the obvious embedding via the 
Lie algebra generators, 
we find novel types of solutions in Section \ref{sec:M-M-embeddings}, 
which are interpreted as momentum and phase space embeddings.
Moreover, in Section \ref{sec:eff_metric_4d_physics} we elaborate on the 
effective metric on these bundles arising from the different embeddings.
Lastly, in Appendix \ref{app:fuzzy_algebra} we provide more technical details of 
the fuzzy geometry.

The paper is intended as a technical resource for  
research involving the $\cS^4_\L$ spheres and similar spaces.
The contents presented should provide all the necessary geometrical and 
algebraic tools for justifying, for example, dimensional reduction, and for 
extracting the low-energy physics on such a background.
However, since we cannot determine the relevant scale parameters of the 
embeddings at this point,
the physical perspectives of the resulting higher-spin theory are only briefly discussed here, and postponed to future work.

\section{Covariant fuzzy four-spheres}
\label{sec:fuzzy_4-sphere}

We are interested in fuzzy 4-spheres which are covariant under $SO(5)$.
They will be defined in terms of five hermitian matrices 
$X^a, \ a=1,\ldots, 5$ acting on some finite-dimensional Hilbert space $\cH$, 
and transforming as vectors under $SO(5)$ 
\begin{align}
 [\cM_{ab},X_c] &= i(\d_{ac} X_b - \d_{bc} X_a), \nn\\
  [\cM_{ab},\cM_{cd}] &=i(\d_{ac}\cM_{bd} - \d_{ad}\cM_{bc} - \d_{bc}\cM_{ad} + 
\d_{bd}\cM_{ac}) \ .
 \label{M-M-relations}
\end{align}
Throughout this paper, indices are raised and lowered with  $g_{ab} = \d_{ab}$.
The $\cM_{ab}=- \cM_{ba}$ for $a,b \in\{1,\ldots, 5\}$ generate a suitable (not 
necessarily irreducible) 
representation of $\mso(5)$ on $\cH$, and $X^a \in \End(\cH)$ are 
operators
 interpreted as quantized embedding functions 
$X^a \sim x^a: \ S^4 \hookrightarrow \R^5$. Then the radius 
\begin{align}
 X^a X_a &= \cR^2 
\end{align}
is a scalar operator of dimension $L^2$. 
We denote the commutator of the $X^a$  by 
\begin{align}
  [X^a,X^b]  &=: i \Theta^{ab} \ .
\end{align}
Such relations constitute a {\em covariant quantum 4-sphere}.

Particular realizations of 
such fuzzy 4-spheres are obtained from generators $\cM^{ab}, \ a,b = 
1,\ldots,6$ of 
$\mso(6) \cong \msu(4)$ via
\begin{align}
 X^a &= r \cM^{a6}, \qquad a = 1,\ldots,5 \ , \qquad \Theta^{ab} = r^2 \cM^{ab} 
\ .
 \label{X-def}
\end{align}
Here $r$ is a scale parameter of dimension $L$, and $\cH$ is some irreducible 
representation (irrep) of $\mso(6)$. 
In much of the paper we will set $r=1$ for simplicity.
The $\mso(5)\subset \mso(6)$ subalgebra is recovered
by restricting the indices of $\cM^{ab}$ to $a,b=1,\ldots,5$.

This class of quantum spheres was considered in \cite{Steinacker:2016vgf} as a 
promising basis for 
a higher-spin theory including gravity. Here we study their fuzzy geometry in 
more detail, 
and provide new embeddings in matrix models, which resolve some of the internal 
structure.

The covariant quantum 4-spheres can be viewed as compact versions of Snyder 
space \cite{Snyder:1946qz,Yang:1947ud}.
The crucial feature is that the classical isometry group (here $SO(5)$) is fully 
realized.
This is in marked contrast to the basic quantum spaces such as the Moyal-Weyl 
quantum plane $\R^4_\theta$,
where the Poisson tensor $\theta^{ab}$ breaks this symmetry.
The price to pay is that the algebra of ``coordinates''  $X^a$  does not close, 
but involves extra generators $\Theta^{ab}$. 
Nevertheless, their proper geometric interpretation allows to proceed with the 
construction 
of physical theories on such spaces via matrix models, leading to fully 
covariant higher-spin theories
with large gauge symmetry, including a gauged version of $SO(5)$.

\subsection{The basic fuzzy 4-sphere revisited}
The simplest example of the above construction is the basic fuzzy 
4-sphere $\cS^4_N$ \cite{Grosse:1996mz,Castelino:1997rv,Ramgoolam:2001zx},
which is obtained for the highest weight irrep  $\cH = \cH_\L$ of $\mso(6)$ with 
$\L = (N,0,0)$.
Throughout this paper we  label  highest weights by their Dynkin indices. 
Using an explicit oscillator realization and/or some  group theory,
one  derives the following relations  
\cite{Castelino:1997rv,Ramgoolam:2001zx,Kimura:2002nq,Steinacker:2016vgf}
\begin{equation}
 \label{XX-R-const}
\begin{aligned}
  X^a X_a &= \cR^2 = R_N^2 \one\, , \qquad  R_N^2 = \frac{1}{4} N(N+4) \; , \\
 \{X_a,\cM^{ab}\}_+ &=  0  \; ,\\
 \frac 12\{\Theta^{ab},\Theta^{a'c}\}_+ \ g_{aa'} &=  \cR^2 \left(g^{bc} -\frac 
1{2\cR^2} \{X^b, X^c\}_+\right) \; , \\
 \epsilon^{abcde} \Theta_{cd} X_e &= (N+2) \Theta^{ab} \, ,
\end{aligned}
\end{equation}
for indices $a,b,\ldots =1,\ldots,5$.
Here $\{\cdot,\cdot\}_+$ denotes the anti-commutator.
The first relation expresses the remarkable fact that $\cH_\L$ remains 
irreducible as 
representation of $\mso(5) \subset \mso(6)$. This is no longer true for generic 
$\L$.
\paragraph{Semi-classical limit and coadjoint orbits.}
To  understand the geometrical meaning of $\Theta^{ab}$, 
it is best to view the fuzzy sphere as 
quantization of the 6-dimensional coadjoint orbit $\C P^3$ of $SO(6)$; this 
point of view 
will naturally carry over to the generalized spheres.
The general construction  is as follows\footnote{See e.g. \cite{Hawkins:1997gj} 
for a nice introduction to  (quantized) coadjoint orbits.}: For any 
given (finite-dimensional) irrep $\cH_\L$ of $SO(6)$, the generators $\cM^{ab} 
\in \End(\cH_\L)$ 
of its Lie algebra $\mso(6)$
are viewed as quantization of the embedding functions
\begin{align}
 \cM^{ab} \sim m^{ab}: \quad \cO_\L \hookrightarrow \R^{15} \cong \mso(6) \ 
\end{align}
of the homogeneous space (coadjoint\footnote{For simplicity we identify the Lie 
algebra with its dual.} orbit)
\begin{align}
 \cO_\L = \{g\cdot \L \cdot g^{-1}; \ g\in SO(6)\} \ \cong SO(6)/\cK  \subset 
\R^{15}\ .
\end{align}
in $\R^{15}$. Here $\cK$ is the stabilizer of $\L$.
The weight $\L \in \mh^* \leftrightarrow H_\L \in \mh$ can be identified with a 
Cartan generator via the Killing form, i.e.  
$\mh^* \cong \mh\subset \mso(6)$ .
This amounts to a quantization of the symplectic manifold $\cO_\L$, in complete 
analogy 
to the quantization of phase space in quantum mechanics. The underlying Poisson 
structure 
on $\cO_\L$ is given by the Kirillov-Kostant symplectic form
\begin{align}
 \{m^{ab},m^{cd}\} = g^{ac}m^{bd}  - g^{ad}m^{bc} - g^{bc}m^{ad} + g^{bd}m^{ac} 
\ , \qquad a,b=1,\ldots,6
 \label{Poisson-m}
\end{align}
whose quantization is the Lie algebra $\mso(6)$. 
Among the 15 functions $m^{ab}$, we use only the 5 functions
\begin{align}
 X^a = \cM^{a6} \sim x^a &= r\, m^{a6}, 
 \end{align}
 which satisfy
 \begin{align}
 \{x^a,x^b\} &= m^{ab}  .
\end{align}
For $\L=(N,0,0)$, there are additional relations analogous to  
\eqref{XX-R-const} such as
  \begin{align}
 x^a x_a &= \cR^2 = \frac {N^2}{4},  \qquad m^{ab} x_b = 0, 
 \qquad  \epsilon^{abcde} m_{cd} x_e = N m^{ab} \ 
 \label{fuzzyS4-class}
\end{align}
for $ a,b,\ldots =1,\ldots,5$,
and we will see that the map
\begin{align}
  x^a: \ \ \cO_\L \cong  \C P^3  \to  S^4\hookrightarrow \R^5 \ 
 \end{align}
is nothing but the Hopf map.
 Hence, fuzzy $S^4$ arises as projection of $\C P^3$ to $\R^5$, and
$\Theta^{ab}$ is the quantization of the Poisson tensor $\theta^{ab}$ on $\C 
P^3$.
We now elaborate this from two different points of view, using 
$\mso(6)\cong \msu(4)$.

\paragraph{$SU(4)$ point of view.}
To see that $\cO_\L \cong \C P^3$ for the basic fuzzy 4-sphere $\cS^4_N$, we 
can view it as $SU(4)$ conjugacy class of
\begin{align}
N H_{\L_1} = \frac N4 \diag(3,-1,-1,-1)  
 = N|\psi_0\rangle\langle \psi_0| -\frac{N}{4} \one_4 , \qquad  \ \ 
|\psi_0\rangle =
 \begin{pmatrix}
1\\0\\0\\0 
\end{pmatrix}
\in \C^4 \ . \nn
\end{align}
The stabilizer of $\L= N \L_1$ is
$\cK=SU(3)\times U(1)$, and clearly $\cO_\L \cong \C P^3$. 
The functions $m^{ab}$ are obtained as 
\begin{align}
 m^{ab} &= \Tr(\xi\, \Sigma^{ab}) , \qquad a,b=1,\ldots,6 , \qquad \xi \in \ 
\cO_\L \; ,
  \label{mab-def-class}
\end{align}
where $\Sigma^{ab}$ are in the 4-dimensional representation of $\mso(6)$.
Using $\Sigma^{a6} = -\frac 12 \g^a$ as in 
\cite{Ho:2001as,Castelino:1997rv,Steinacker:2016vgf}, 
the reference point 
$\xi_0 = H_\L$ in $\C P^3$ is mapped via $x^a$ to the ''north pole`` 
$p=(0,0,0,0,N)$ of $S^4 \subset  \R^5$,
and the fiber over $p$ is recognized as a 2-sphere $S^2$.
Thus, $\C P^3$ is a $S^2$ bundle over $S^4$. The $SO(5)$ rotations are obtained 
from its spinorial representation on $\C^4$.
\paragraph{$SO(6)$ point of view.}
Alternatively, we can view  $\cO_\L$ as  $SO(6)$ orbit. Then the $SO(5)$ 
symmetry is manifest, which is important to understand the generalized spheres.
To identify this orbit, we need the 6-dimensional representation of the 
fundamental weights (or rather their duals) 
$H_{\L_1} = \frac 12(\cM_{12}+\cM_{34} + \cM_{56})$ and 
$H_{\L_3} = \frac 12(-\cM_{12}+\cM_{34} + \cM_{56})$ and 
$H_{\L_2} = \cM_{56}$. Since  $\pi_6^{ij}(\cM_{ab}) = \d^i_a\d^j_b - 
\d^i_b\d^j_a$, these are
\begin{equation}
\begin{aligned}
 H_{\L_1} &= \frac 12 \begin{pmatrix}
       & 1 &&&&\\
       -1 &&&&&\\
       &&& 1 &&\\
       &&-1&&& \\
       &&&&& 1\\
       &&&&-1&
      \end{pmatrix}, \quad 
  H_{\L_2} = \begin{pmatrix}
       & 0 &&&&\\
       0 &&&&&\\
       &&& 0 &&\\
       && 0&&& \\
       &&&&& 1\\
       &&&& -1&
      \end{pmatrix} ,  \\ 
  H_{\L_3} &= \frac 12\begin{pmatrix}
       & 1 &&&&\\
       -1 &&&&&\\
       &&& -1 &&\\
       && 1&&& \\
       &&&&& 1\\
       &&&& -1&
      \end{pmatrix}  \ .
\end{aligned}
\label{fund-weights-so6}
\end{equation}
To recognize $\cO_\L = \cO_{N\L_1} \cong\C P^3$ as  $SO(6)$ orbit, we note that 
the stabilizer group of $\L_1$
is given by $\cK = SU(3) \times U(1)$, which is realized as 
\begin{align}
 U = \begin{pmatrix}
      a&b&c \\
      d&e&f \\
      g&h&i
     \end{pmatrix} \in   U(3) \subset SO(6) \ .
  \label{SU3-irreg}
\end{align}
Here the complex numbers are identified with $2\times 2$ matrices as $a = 
\a\one_2+\b (i\s_2) \cong \a+i\b\in \C$ etc.,
identifying   $\scriptsize\begin{pmatrix}
                                                                               0 
& 1\\ -1& 0
                                                                              
\end{pmatrix} = i\s_2$ with $i$.
We also note that the Weyl group $S_3$ of this $SU(3)$
acts by permuting these $2\times 2$ blocks.                       
In particular, the embedding functions $m^{ab}$ of \eqref{mab-def-class} are 
now given by
\begin{align}
 m^{ab}: \quad \cO_\L &\hookrightarrow \R^{15} \ \nn\\
    \xi &\mapsto  (\xi)^{ab}  \; ,
 \label{cM-def-CP3}
\end{align} 
 which satisfies the characteristic matrix equation 
\begin{align}
 \m^2 = -\frac{N^2}{4}, \qquad \mbox{i.e.} \quad  m^{ac} m^{cb} = -\frac{N^2}{4} 
\ g^{ab} \ .
 \label{char-CP3-class}
\end{align}
Here and in the following we use boldface-notation
\begin{align}
 \m = (m^{ab})
\end{align}
to indicate matrices of functions.
The projection to $S^4$ is given by
$x^a = m^{a6}$ i.e. by the 6th column of the $6\times 6$ matrices 
$\xi\in\cO_\L$.
Using  \eqref{fund-weights-so6}, we see that
the reference point $\xi_0 = H_{N\L_1}$ in $\C P^3$ is projected to the ''north 
pole`` $p=(0,0,0,0,N)$ of $S^4 \subset  \R^5$,
whose stabilizer $SO(4)\subset SO(6)$  is generated by
the $\cM^{\mu\nu}$ with  $\mu,\nu \leq 4$. 

Now we come to a very important point. 
This $SO(4) \sim SU(2)_L \times SU(2)_R$ stabilizer of the north pole $p \in 
S^4$ 
acts non-trivially on the fiber in $\C P^3$ over $p$.
More precisely, $SU(2)_L$ generates an $S^2$ orbit of self-dual $4\times 4$ 
matrices through $\diag(i\s_2,i\s_2)$,
while $SU(2)_R$ acts trivially.
To indicate this action of the local isometry group on the fiber, we say that 
{\bf $\C P^3$ is a twisted bundle over $S^4$}; more precisely it is an $SO(5)$ 
equivariant bundle 
\cite{citeulike:12335164}. 
In contrast to the 
conventional Kaluza-Klein compactifications\footnote{A somewhat analogous 
structure is realized in 
twisted supersymmetry, cf \cite{Vafa:1994tf}.}, the harmonics on $S^2$ then 
transform non-trivially and lead to  
higher spin fields. 
Explicitly, the functions on $\cO_\L \cong \C P^3_N$ decompose into the direct 
sum of 
higher spin harmonics on $S^4$  \cite{Steinacker:2015dra}
\begin{align}
 \cC(\C P^3) =  \bigoplus\limits_{0 \leq n} (n,0,n)_{\mso(6)} 
 \cong \bigoplus\limits_{0 \leq m\leq n} (n-m,2m)_{\mso(5)}  \ .
 \label{mode-expansion-alg}
\end{align}
This is the twisted analog of a KK tower.
For example, $m^{ab}$ lives in $(1,0,1) = (1,0) \oplus (0,2)$ and decomposes 
into a scalar function and a 2-form on $S^4$.

Now consider the Poisson structure on $\C P^3$. 
Its projection (push-forward) to the base $S^4$ defines a bundle of bi-vectors 
\begin{align}
 \theta^{\mu\nu}(x,\xi)\del_\mu \otimes \del_\nu \ 
 \end{align}
over $S^4$.
Here $\xi$ are coordinates on the internal $S^2$ fiber of $\cO_\L$ over $S^4$,
and  $\del_\mu$ are tangent vectors of $S^4$.
Due to \eqref{fuzzyS4-class}, $\theta^{\mu\nu}(x,\xi)$ is  self-dual  at each 
$x\in S^4$.
This defines a bundle of self-dual 2-forms over $S^4$,
which transform along the fiber $S^2$ as $(1,0)$ under the local $SO(4) = 
SU(2)_L\otimes SU(2)_R$ via 
$\{\theta^{\mu\nu},.\}$. In the non-commutative case, this amounts to a gauge 
transformation
\begin{align}
\L^{\mu\mu'}  \L^{\nu\nu'} \Theta^{\mu'\nu'}
 = U_\L\Theta^{\mu\nu}U_\L^{-1} \ .
\end{align}
In other words, local rotations are implemented as gauge transformations, as 
desired in gravity. 
This provides a covariant type of non-commutative geometry, by ''averaging``   
 $\theta^{\mu\nu}$ (i.e. the $B$-field in string language) at each point of the 
$S^4$ space.
\section{Twisted bundles over \texorpdfstring{$S^4$}{S4} from coadjoint 
\texorpdfstring{$SO(6)$}{SO(6)} orbits}
\label{sec:twisted_bdl_coadj_orbit}
Armed with these insights, we can proceed to the generalized fuzzy spheres 
$\cS^4_\L$. They are defined by the same relations \eqref{M-M-relations} 
through \eqref{X-def} as the basic $S^4_N$, 
where $\cM^{ab}$
are now the generators of the irreducible representation with 
highest weight\footnote{The most general case $\L=(N,m,n)$ is postponed for 
future work.}
$\L=(N,0,n)$. 
The underlying classical geometry is then the 10-dimensional coadjoint orbit 
$\cO(\L)$ of $SO(6)$.
For $N\gg n$,
they can be viewed either as twisted bundle over fuzzy $\C P^3$ with fiber $\C 
P^2$, or 
as twisted bundle over $S^4_N$ or  $S^4_n$.
The word ''twisted`` again indicates a non-trivial action of the local isometry 
group on the fiber.
We will identify several embeddings of this space into target space which make 
this structure manifest,
and which provide the classical analogs of the matrix model solutions discussed 
in Section \ref{sec:M-M-embeddings}.
\subsection{Classical geometry}
To understand the geometry of  $\cO_\L$ for $\L=(N,0,n)$, we can view it either 
as $SU(4)$ orbit or as $SO(6)$ orbit.
\paragraph{$SU(4)$ point of view.}
We first view  $\cO_\L$ as $SU(4)$ conjugacy class of $H_\L = N H_{\L_1} + n 
H_{\L_3}$, where
\begin{equation}
\begin{aligned}
H_{\L_1} &= \frac 14 \diag(3,-1,-1,-1) \,, 
\qquad   H_{\L_2} = \frac 12 \diag(1,1,-1,-1)\, , \\
  H_{\L_3} &= \frac 14 \diag(1,1,1,-3) \, . 
\end{aligned}
\end{equation}
Thus 
\begin{align}
 H_\L = \frac 14\begin{pmatrix}
         3N+n & & & \\
         & -N+n & & \\
          & & -N+n & \\
          & & & N-3n
              \end{pmatrix},
\end{align}
with stabilizer 
$SU(2)\times U(1)\times U(1)$. Hence, $\cO_\L$ is the 10-dimensional
manifold of traceless $4\times 4$ hermitian matrices $\xi$ which satisfy 
the characteristic equation 
\begin{align}
 \left(\xi- \frac{3N+n}4 \right) 
 \left(\xi- \frac{-N+n}4 \right)
 \left(\xi - \frac{-N-3n}4 \right) = 0 \ .
\end{align}  
If we assume $N \gg n$, then one eigenvalue $\approx \frac{3N}4$ is large and 
the other 3 eigenvalues $\approx -\frac{N}{4}$
approximately coincide, almost as in $\C P^3$. 
In fact, $\cO_\L$ is naturally a bundle over $\C P^3$,
\begin{align}
 \xymatrix{\mathrm{CP}^2 \ar@{^{(}->}[r] & \cO_{\Lambda} \ar[d]_{P_N} \\
 & \mathbb{CP}^3  \cong\cO_{N \Lambda_1} \ar[d]^{x^a} \\
& S^4  
}
\end{align}
where $P_N$ is a spectral map (i.e. some polynomial) which maps $H_{\L}$ to 
$H_{N\L_1}$.
Similarly, we define $P_n$ to be a spectral map which maps $H_{\L}$ to 
$H_{n\L_3}$.
For a point $\xi \in \cO_\L$, the fiber $P_N^{-1}(e)$ over $e=P_N(\xi)\in 
\cO_{N 
\L_1}$ is given by
\begin{align}
 P_N^{-1}(e)
 = \left\{ \mathrm{Ad}_h(\xi) \, | \, h \in \mathrm{Stab}_{SU(4)}(e) \right\} 
=e+ \left\{\mathrm{Ad}_h(P_n(\xi)) \, | \, h \in \mathrm{Stab}_{SU(4)}(e)  
\right\} \; ,
\end{align}
where we used $\xi = P_N (\xi) +P_n (\xi)$.
Since the stabilizer of $H_{\L_1}$ is $SU(3)$,
the fiber $P_N^{-1}(e)$ is given by the (shifted) $SU(3)$ coadjoint orbit
of the remaining matrix $P_n(\xi)$, which is $\C P^2$.
Hence, $\cO_\L$ is a $\C P^2$ bundle over $\C P^3$.

Of course the same story applies if we interchange $N$ with $n$, so that we 
have 
the following picture:
\begin{align}
 \xymatrix{ & \cO_{\Lambda} \ar[dl]_{P_N} \ar[dr]^{P_n} & \\
 \mathbb{CP}^3  \cong\cO_{N \Lambda_1} \ar[d]^{x_N^a} &  & \cO_{n \Lambda_3} 
\cong \mathbb{CP}^3 \ar[d]^{x_n^a} \\
S^4_N & & S^4_n
}
\label{S4-S4-bundle-proj}
\end{align}
%
%
\paragraph{$SO(6)$ point of view.}
To make the 4-sphere and its $SO(5)$ symmetry manifest, it is better to view 
$\cO_\L$
as $SO(6)$ orbit generated by \eqref{fund-weights-so6}
\begin{align}
 H_\L = NH_{\L_1}+ n H_{\L_3} =
 \frac 12\begin{pmatrix}
       & N+n &&&&\\
       -N-n&&&&&\\
       &&&N-n&&\\
       &&-N+n&&& \\
       &&&&& N+n\\
       &&&&-N-n&
      \end{pmatrix} 
 \label{generalized-sphere-so6-orbit}
\end{align}
which provides the embedding functions $m^{ab}$ of \eqref{mab-def-class} as 
follows:
\begin{align}
 m^{ab}: \quad \cO_\L &\hookrightarrow \R^{15} \ \nn\\
    \xi &\mapsto  (\xi)^{ab} 
 \label{cM-def-full}
\end{align} 
as anti-symmetric $6\times 6$ matrices.
This allows to construct explicit embeddings of $\cO_\L$ into some 
lower-dimensional target space $\R^D$, 
which realize the above bundle projections 
$\cO_\L \stackrel{P_{N,n}}{\rightarrow} \C P^3 \rightarrow S^4_{N,n}$ in a 
$SO(5)$-covariant way.
We will also provide an explicit embedding of the internal $\C P^2$ fiber.
This is the basis for the matrix embeddings of their fuzzy counterparts.
\paragraph{$S^4 \times S^4$ structure.}
As indicated above, the projection to $\C P^3$ can be realized by a polynomial 
map which appropriately changes the eigenvalues.
This is achieved by the following matrix-valued functions on $\cO_\L$:
\begin{align}
 \begin{pmatrix}
  m_N^{ab} \\ m_n^{ab} 
 \end{pmatrix}
&= A^{(c)} \begin{pmatrix}
  \tilde m^{ab} \\ m^{ab} 
 \end{pmatrix}, \nn\\[1ex]
 \quad A^{(c)} &=  \frac{1}{2(N^2-n^2)}\begin{pmatrix}
                                     4 & 3N^2+n^2 \\
                                     -4 & - (N^2+3 n^2)
                                    \end{pmatrix}
 \eqqcolon \begin{pmatrix} 
 \alpha_N^{(c)} & \beta_N^{(c)} \\  \alpha_n^{(c)} & \beta_n^{(c)} 
\end{pmatrix}
 \label{MNn-trafo-class}
\end{align}
(the superscript ${(c)}$ stands for classical), where
\begin{align}
 \tilde m^{ab} 
= (\m^3)^{ab} \equiv m^{ac}  m^{cd}  m^{db} = - \tilde m^{ba} \ .
\end{align}
The original embedding \eqref{cM-def-full} of $\cO_\L$ is 
recovered as 
\begin{align}
 m^{ab} = m^{ab}_N + m^{ab}_n \ .
 \label{MisMNplusMn}
\end{align}
Both $m_N^{ab}$ and  $m_n^{ab}$ are anti-symmetric
and can, hence, be viewed as elements of $\mso(6)$.
They are constructed such that their eigenvalues are $\pm i N$ and $\pm i n$, 
respectively, and 
\begin{align}
  m_N ^{ab} = (O H_{N\L_1} O^{T}) ^{ab}, \qquad  m_n^{ab} = (O H_{n\L_3} O^{T}) 
^{ab} \ ,   \qquad O \in SO(6) \ .
\end{align}
Recalling the discussion of $\cO_{\L_1}$, 
this means that they describe $\C P^3$ with ''radius`` $N$ and $n$, 
respectively. 
Therefore, $m_{N,n}^{ab}$ realize the two bundle projections 
\eqref{S4-S4-bundle-proj} to $\C P^3_N$ and $\C P^3_n$, with
 stabilizer groups $\cK = SU(3)\subset SO(6)$ as defined in \eqref{SU3-irreg}.
The further projections to $S^4_N$ and $S^4_n$ can then be defined as before
\begin{align}
 x_N^a &=  m_N ^{a6}, \qquad \  x_n^a =  m_n ^{a6}, \qquad a=1,\ldots,5 \, ,
\end{align}
and the original embedding  $x^a: \ \cO_\L \to \R^5$ of the generalized 
4-sphere 
\eqref{MisMNplusMn} is recovered as
\begin{align}
 x^a = x_N^a + x_n^a \,.
 \label{eq:XN+Xn_classical}
\end{align}
All the identities \eqref{fuzzyS4-class}, including the self-duality 
statements, hold for both $x_N$ and $x_n$,
which realize semi-classical $S^4_N$ and $S^4_n$, with  inner products
\begin{subequations}
\label{xNn-matrix-class}   
\begin{align}
 \begin{pmatrix}
  x_N\cdot x_N & x_N\cdot x_n \\
   x_n\cdot x_N & x_n\cdot x_n  
 \end{pmatrix}
&= 
\begin{pmatrix}
 {R_{N}^{(c)}}^2 & 0 \\
 0 & {R_{n}^{(c)}}^2
\end{pmatrix}
 + \delta \begin{pmatrix}
        0 & 1\\
        1 & 0
       \end{pmatrix},
\end{align}
 where     
 \begin{equation}
  \begin{aligned}      
\delta &\coloneqq x_N  \cdot x_n =  \frac{1}{2} \left(\cR^2 -  
\left({R_{N}^{(c)}}^2 +{R_{n}^{(c)}}^2\right)\right) ,  \\
{R_{N}^{(c)}}^2 &= \frac 14 N^2, \qquad {R_{n}^{(c)}}^2 = \frac 14 n^2 \ .
\end{aligned}
\end{equation}
\end{subequations}
Here $x_N \cdot x_N \equiv x_N^a {x_N}_a$ etc.
This means that  $S^4_\L$ can be viewed as a sum of two basic 4-spheres 
$S^4_N$ and $S^4_n$ with 
radii $R_{N,n}^{(c)}$. This explains why its ''radius``
\begin{align}
 \cR^2 = x\cdot x  \neq \mathrm{const}
\end{align}
is a non-trivial function on $\cO_\L$ taking values in the interval
\begin{equation}
 \cR \in   [r_{\rm min},r_{\rm max}]  =  
\left[\frac{N-n}{2},\frac{N+n}{2}\right]\,
\label{eq:intercal_cR}
\end{equation}
where the extremal values correspond to parallel and anti-parallel $x_N$ and 
$x_n$, respectively.
The same will hold in the fuzzy case.

Besides $m_N^{ab}$ and $m_N^{ab}$, there are no further anti-symmetric 
matrix-valued functions in $(1,0,1)$, 
because the multiplicity of these modes in the (polynomial) algebra 
of functions on $\cO_\L$ is two\footnote{For generic $\L=(N,n_1, n_2)$ there 
would be 3 such functions, including $(m^5)^{ab}$.}.
We also observe that $(\alpha_N^{(c)}, \beta_N^{(c)}) \ \leftrightarrow \ 
(\alpha_n^{(c)}, \beta_n^{(c)})$
upon interchanging $N \leftrightarrow n$. This reflects the outer automorphism 
$\Omega$  of $\mso(6)$,
replacing $\L_1 \leftrightarrow \L_3$.
\paragraph{Squared orbit and $\C P^2$ fiber.}
Similarly, the fibers of the above projections $\cO_\L \to \C P^3$ can be 
captured by the 
 following matrix-valued function on $\cO_\L$:
 \begin{equation}
 \begin{aligned}
  t^{ab} &\coloneqq -(\m^2)^{ab}  = t^{ba} \, , \\
  t^a  &\coloneqq t^{a6}, \qquad \  t^6 = t^{66} =  \cR^2 \,,
 \end{aligned}
 \end{equation}
where 
\begin{align}
 \Tr (\t) = - \Tr(\m^2) = \frac 12(3N^2+3n^2+ 2nN)
\end{align}
is (twice) the classical quadratic Casimir of $\mso(6)$.
The $t^{ab}$ are symmetric  matrices which describe 
the conjugacy class of 
\begin{align}
 - H_\L^2  &=  \frac 14 (N^2+ n^2)\one_6 + \frac 12 nN 
\diag(\one_2,-\one_2,\one_2)   \ .
\end{align}
It is convenient to consider instead the shifted matrix
\begin{align}
 \tilde t^{ab} = t^{ab} - \big({R_{N}^{(c)}}^2 +{R_{n}^{(c)}}^2\big)  g^{ab} 
\qquad 
 \mbox{such that}  \ \     \tilde t^{66} = 2\delta= 2x_N\cdot x_n\ .
 \label{shifted-t}
\end{align}
They describe the $SO(6)$ orbit 
\begin{align}
 \tilde t^{ab} = \frac 12 Nn\, (O \diag(\one_2,-\one_2,\one_2)O^T)^{ab} \ , 
\qquad O \in SO(6)
 \label{O-orthogonal}
\end{align}
which is an 8-dimensional manifold 
of orthogonal $6\times 6$ matrices rescaled by $\frac 12 Nn$.
However, as a $\cK=SU(3)$ orbit under the stabilizer $\cK$ of either $\C P^3$ 
base, 
it is nothing but $\C P^2$. This can be seen by using the identification of 
$\C$ 
in \eqref{SU3-irreg} ff.,  noting that 
the $SU(3)$ orbit through $\diag(1,-1,1)$ is $ SU(3)/SU(2)\times U(1) \cong \C 
P^2$.
We recover the fact that  $\cO_\L$ is a $\C P^2$-bundle over $\C P^3$, and $t^a$ 
provides a $SO(5)$-covariant 
embedding of $\C P^2 \to \R^5$. This embedding has interesting degeneracies, 
which will be exhibited in the next section.
\paragraph{Matrix algebra.}
Besides $t^{ab}$, there are no further symmetric matrix-valued functions in 
$(0,2,0)$ in the  algebra 
of functions on $\cO_\L$. 
Together with the corresponding statement for $m_{n,N}^{ab}$, this means that 
  $m_N^{ab}, m_N^{ab}, t^{ab}$, and $g^{ab}$ 
form a closed algebra under matrix multiplication, which is encoded in the 
characteristic equation 
\begin{align}
 \Big(\m^2 + \frac{(N-n)^2}{4}\Big)\Big(\m^2 + \frac{(N+n)^2}{4}\Big) = 0 
 \label{char-generic-class}
\end{align}
i.e. 
\begin{align}
\m^4 = \t^2 = \frac 12(N^2+n^2) \t - \frac 1{16}(N^2-n^2)^2 \ .
\label{char-generic-class-2}
\end{align}
Observe that $\t$ satisfies a quadratic equation, reflecting its relation to $\C 
P^2$.
Rewriting this matrix algebra in terms of $m_{N,n}^{ab}$ and $\tilde t^{ab}$,
we obtain
\begin{equation}
\label{MNnO-algebra-class}
 \begin{aligned}
 \m_N \cdot \m_N &= - \frac{N^2}{4} g \, , \qquad   \m_n\cdot \m_n = - 
\frac{n^2}{4} g \, ,\\
   \m_N \cdot \m_n &= -\frac 12\, \tilde \t \, , \\
 \tilde \t \cdot \m_N &= \frac{N^2}2\, \m_n \, , \qquad 
 \tilde \t \cdot \m_n =  \frac{n^2}2\,  \m_N \, ,\\
 \tilde \t\cdot \tilde \t &=  \frac {N^2 n^2}4 \ g \ .
\end{aligned}
\end{equation}
Here $\cdot$ indicates matrix multiplication as $6\times 6$ matrix, e.g. $\m_N 
\cdot \m_n = m_N^{ac} m_n^{cb}$.
The algebra is commutative, $\tilde \t\cdot \m_N =  \m_N\cdot\tilde \t$ etc.
The first relations are nothing but the characteristic equation  
\eqref{char-CP3-class} for $\C P^3$.
Taking $a6$ matrix elements, we obtain
\begin{equation}
\label{MNnO-vector-algebra-class}
\begin{aligned}
  \m_N \cdot x_N &= 0 =  \m_n\cdot x_n \\
  \m_N \cdot x_n &= -\frac 12\, \tilde t  =   \m_n \cdot x_N \\
  \m_N \cdot \tilde t &=\frac{N^2}2\, x_n, \qquad  \m_n \cdot \tilde t 
=\frac{n^2}2\, x_N
\end{aligned}
\end{equation}
etc.
Finally,
taking the $66$ matrix element and using anti-symmetry of $m_{N,n}^{ab}$, we 
obtain the inner products of 5-vectors
\begin{equation}
\label{scalar-products-class}
\begin{aligned}
 t\cdot x_n &= 0 =  t\cdot x_N  \\
 x_N \cdot x_n &= \delta  = \frac{1}{2}\left(\cR^2-\frac{1}{4} 
\left(N^2+n^2\right)\right) \\
 \sum_{a=1}^6 \tilde t^a \tilde t_a &= 
 4\d^2 + \sum_{a=1}^5 t^a t_a =  \frac {N^2 n^2}4
\end{aligned}
\end{equation}
using \eqref{shifted-t}.
Note in particular that the 5-vector $t=\tilde t$ is perpendicular to both $x_n$ 
and $x_N$.
Hence, for fixed points $x_N\in S^4_N$ and $x_n\in S^4_n$, $ \ t$ is in the 
tangent plane 
of both spheres, and generically sweeps out an $S^2$ whose radius depends on 
$\cR^2$.
Also, observe that the 5-vector $t$ vanishes $t\cdot t = 0$ if and only if 
$\cR^2$ is extremal,
i.e.\ if $x_N$ and $x_n$ are (anti-)parallel.
\subsection{Global aspects}
\label{sec:global}
\paragraph{$SO(5)$ fibration.}
The functions $x_N^a, x_n^a$, and $t^a$ define a $SO(5)$-covariant map 
\begin{align}
 \xymatrix{\cO_\L  
 \ar[d]_{(x_N, x_n,t)}& \\
 \tilde \cO_{\Lambda} & \subset S^4_N\times S^4_n \times \R^5 \subset \R^{15}
}
\label{O-SSS-map}
\end{align}
Although this map is a local immersion at generic points, we will see that it 
has a non-trivial global structure.
In particular, there is a degenerate 3-fold covering or self-intersection near 
the reference point $\xi = H_\L \in \cO_\L$. 

First, we focus on the embedding $\tilde \cO_{\Lambda}  \subset S^4_N\times 
S^4_n\times \R^5 $,
which is described by the constraints 
\begin{equation}
 \label{constraints-explicit}
\begin{aligned}
 x_N\cdot t &= 0 = x_n \cdot t  , \\
 x_N \cdot x_n &=  \frac{1}{2} \left(\cR^2-\frac{1}{4} (N^2+n^2)\right), \\
 t\cdot t &= \Big(\cR^2 - \frac{1}{4}(N-n)^2\Big)\Big(\frac{1}{4}(N+n)^2 - 
\cR^2 \Big) \,,
\end{aligned}
\end{equation}
which follow from \eqref{scalar-products-class}.
Generically, these are 4 independent conditions, 
which define a 9-dimensional $SO(5)$ orbit $\tilde\cO_\cR$ labeled by $\cR$. 
The last equation reflects the fact that 
$\cR \in \left[\frac{N-n}{2},\frac{N+n}{2}\right]$, see \eqref{eq:intercal_cR},
where the extremal values correspond to parallel and anti-parallel $x_N$ and 
$x_n$, respectively.

Assume first that $\cR^2$ is neither maximal nor minimal, so that
$t\cdot t \neq 0$, and $x_N$ and $x_n$ are not aligned.
Fix e.g. $x_N \sim (0,0,0,0,1) \in S^4_N$. Its stabilizer $SO(4)\subset SO(5)$
acts non-trivially on $t \in \R^4$, which is tangential due to $t\cdot x_N=0$ 
and non-vanishing by assumption.
Hence, $t$ sweeps out some $S^3 \subset \R^4$. Fix again $t_0 \sim (0,0,0,1) 
\in 
S^3$. Its stabilizer $SO(3)\subset SO(4)$ still acts on $x_n$,
which is perpendicular to $t$ and not parallel to $x_N$. Hence, $x_n$ sweeps 
out 
some $S^2$. 
Therefore, $\cO_\cR$ is a homogeneous  $SO(5)$ space with  
local structure\footnote{Note that this is not a $SO(5)$ coadjoint orbit, and 
therefore it can have odd  dimension.}
\begin{align}
 \tilde\cO_\cR \cong S^4\times S^2\times S^3  \ \subset \ S^4_N\times S^4_n 
\times \R^5
\end{align}
described by $(x_N, x_n,t)$. Thus, $\tilde\cO_\L$ decomposes into these $SO(5)$
orbits $\tilde\cO_\cR$ labeled by $\cR\in [r_{\rm min},r_{\rm max}]$, which 
become degenerate 
at the endpoints. 

\begin{figure}[h]
\centering
\begin{subfigure}[b]{0.45\textwidth}
 \includegraphics[width=1\textwidth]{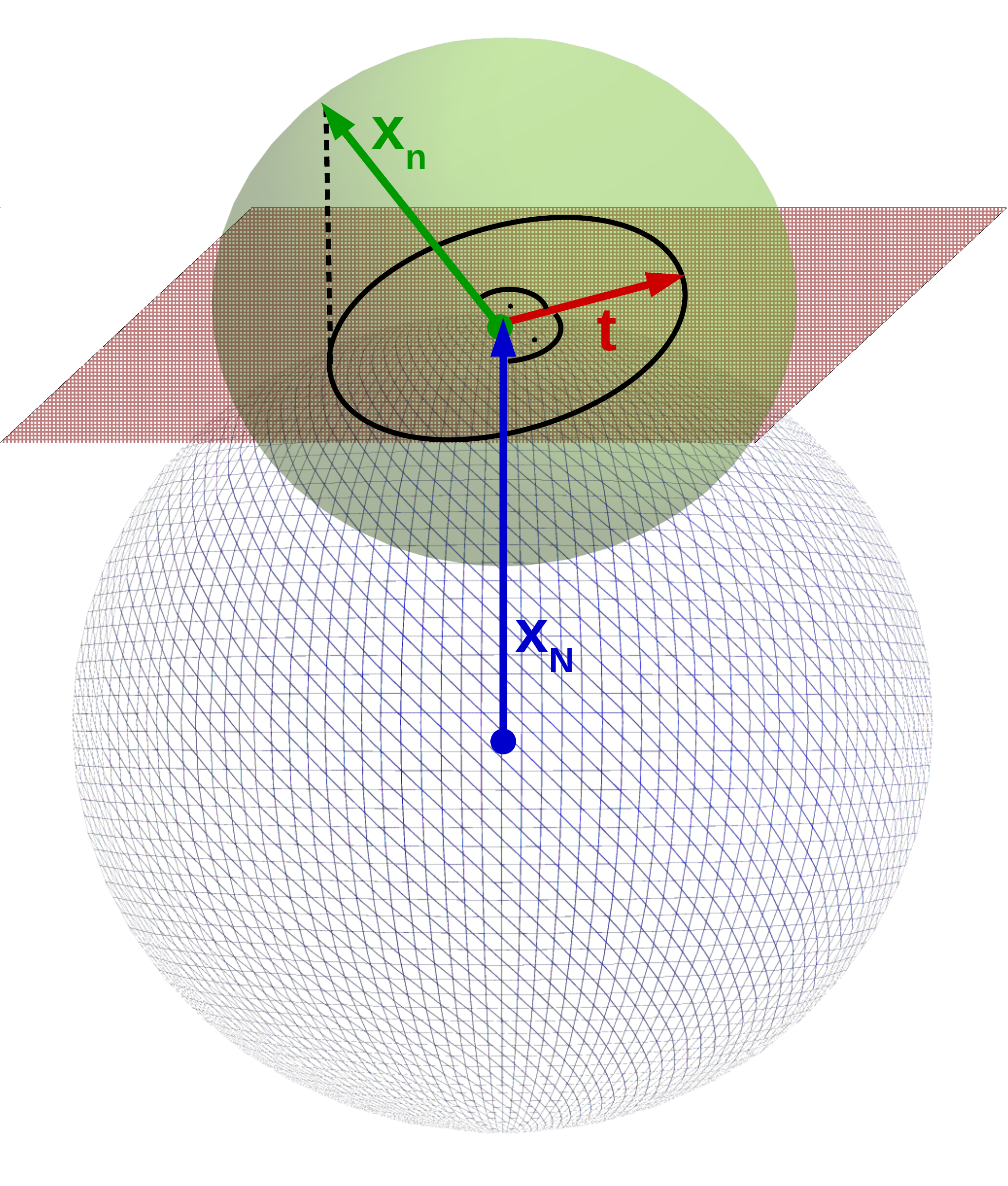} 
 \caption{ }
 \label{fig:gen_coordinates}
\end{subfigure}
\begin{subfigure}[b]{0.45\textwidth}
 \includegraphics[width=1\textwidth]{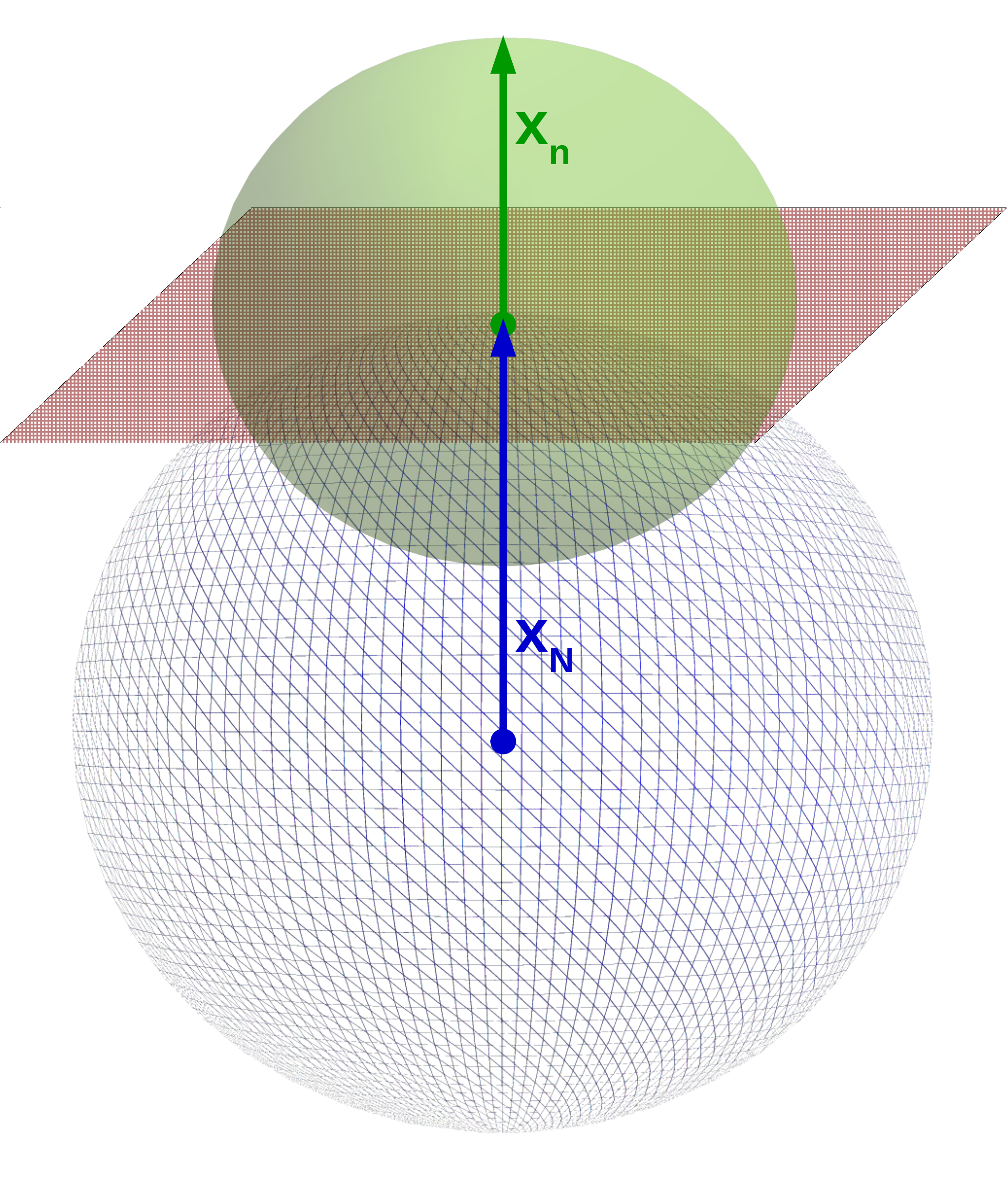} 
 \caption{ }
 \label{fig:extremal_coordinates}
\end{subfigure} 
 \caption{Sketched geometry of the coordinates $(x_N,x_n,t)$ for fixed 
$\cR^2$. (a) For $x_N \nparallel x_n$, $x_N$ sweeps out a large (blue 
shaded) sphere $S^4_N$ and $x_n$ sweeps out a smaller (green) sphere $S_n^4$.  
$t$ lies in the (red) tangent space of $S^4_N$ at a point $x_N$. The projection 
of $x_n$ onto the tangent space yields the sphere $S^2_\cR$.  (b) For $x_N 
\parallel x_n$, the projection of $x_n$ onto the tangent space vanishes, i.e. 
$t$ is not a meaningful coordinate.}
 \label{fig:coordinate_spheres}
\end{figure}

Alternatively, we can pick two linearly independent points  $x_N\in S^4_N$ and 
$x_n\in S^4_n$ and write
\begin{align}
  \tilde\cO_\L \cong S^4_N\times S^4_n\times S^2_\cR \ ;,
  \label{embedding-xNxnt}
\end{align}
see also Figure \ref{fig:gen_coordinates}. However, the radius of $t \in 
S^2_\cR$ is encoded in $x_N\cdot x_n$, and vanishes 
 for extremal $\cR^2$, which is sketched in  Figure 
\ref{fig:extremal_coordinates}.

 Since both $ \tilde\cO_\L$ and $\cO_\L$ are 10-dimensional, this provides also  
a faithful local description of $\cO_\L$
 for generic points.
However, the map \eqref{O-SSS-map} is not injective, and 
it turns out to be a degenerate triple cover at least near $t=0$. 
This will be elaborated next.
\paragraph{$SU(3)$ structure and degeneracy.}
To resolve the extremal values of $\cR^2$ and to exhibit the global structure, 
consider the reference point $\xi = H_\L \in \tilde\cO_\L$. 
This corresponds to $x_N=N e_5, x_n = n e_5$ and $t=0$ as 5-vectors, which 
is a non-generic case in the above description since $\cR^2$ takes its 
maximum value. We will exhibit the triple covering structure over  $(x_N,x_n,t)$ 
near $t=0$. 

\begin{figure}
\centering
\begin{subfigure}[b]{0.40\textwidth}
 \includegraphics[width=1\textwidth]{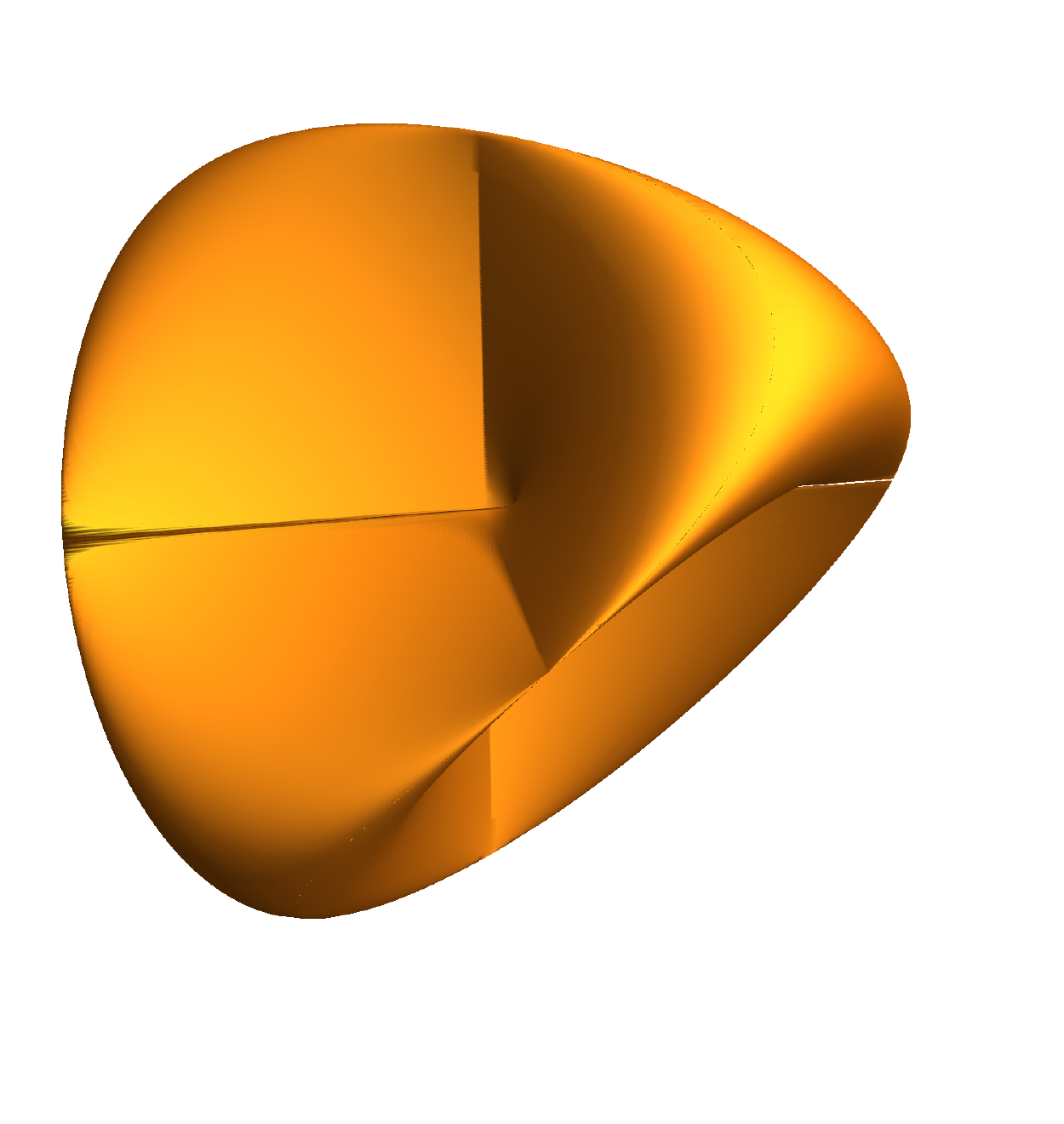} 
 \caption{ }
 \label{fig:squashed-CP2a}
\end{subfigure}
\begin{subfigure}[b]{0.40\textwidth}
 \includegraphics[width=1\textwidth]{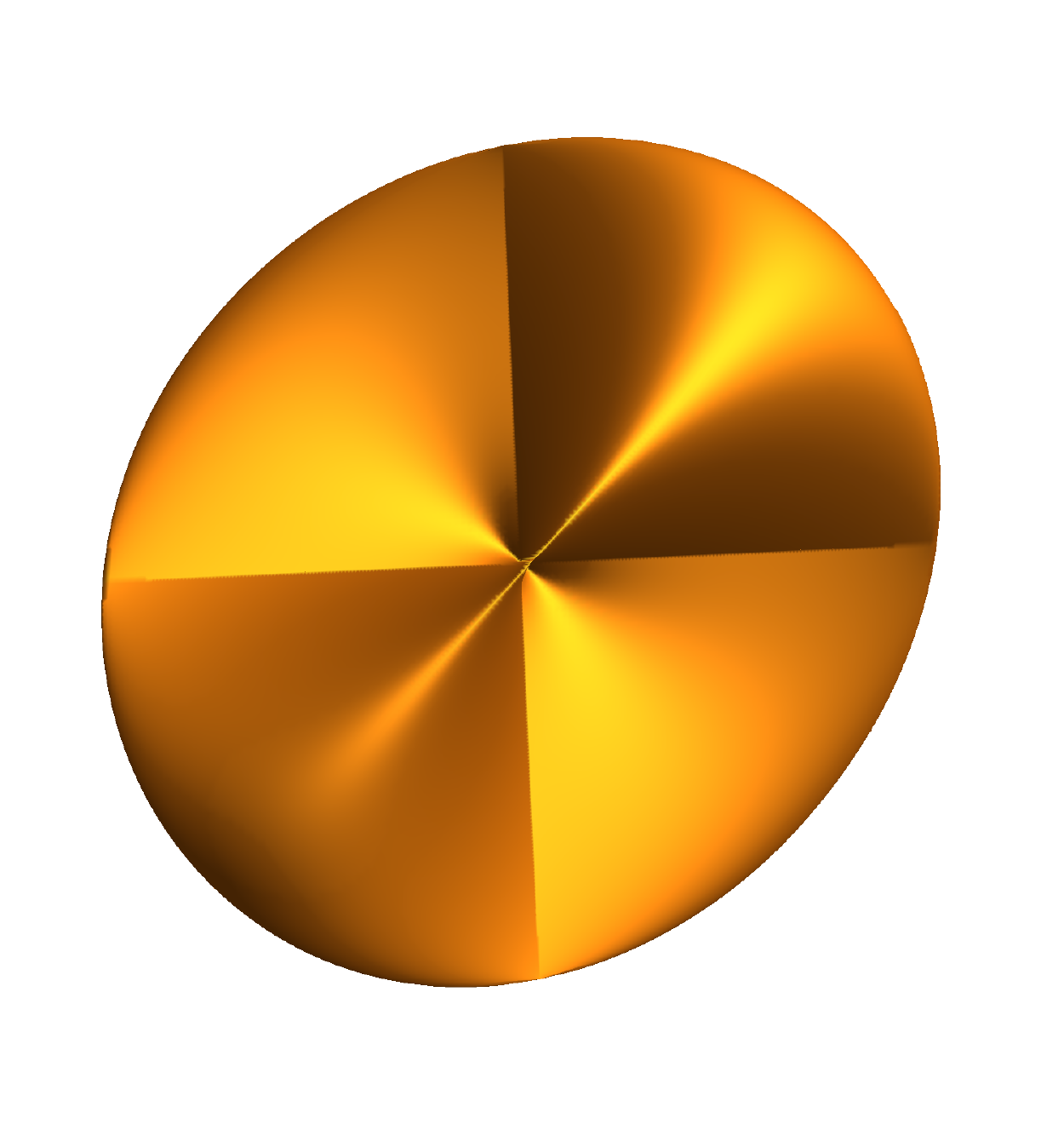} 
 \caption{ }
 \label{fig:squashed-CP2b}
\end{subfigure} 
 \caption{(a) Squashed $\C P^2$ described by 
$(t^1,t^2,t^3,t^4,s^3,s^4)$, with 
triple self-intersection. (b) Squashed $\C P^2$ described 
by $(t^1,t^2,t^3,t^4)$, 
with degenerate covering.}
 \label{fig:squashed-CP2}
\end{figure}

Following the general analysis,  the projection $P_N$ maps $\xi$ to $\xi_N= 
H_{N\L_1}$, which is a point in 
$\C P^3 = \cO_{N\L_1} = SU(4)/SU(3)\times U(1)$ described by
$m_N^{ab}$. Denote with $\cK$ its $SU(3)$ stabilizer, which is  
explicitly given by the $6\times 6$ matrices in \eqref{SU3-irreg}. 
Note that this $SU(3)$ does {\em not} respect $\cR^2$, and it provides the 
missing local parametrization (replacing $t$) of $\cO_\L$
near $\xi$. $\cK$ acts on the symmetric $\tilde t^{ab}$
on $\cO_\L$ via
\begin{align}
  \tilde t^{ab} = \frac 12 Nn\, (U \diag(\one_2,-\one_2,\one_2) U^{-1})^{ab} \ , 
\qquad U \in SU(3) \subset SO(6)
 \label{SU(3)-t}
\end{align}
and similarly on $t^a$. We focus on the linearized action of $U=e^{i\l}$ on 
$\tilde t^{ab}_0 = \diag(\one_2,-\one_2,\one_2)$. 
Since the diagonal (Cartan) generators act trivially on $\tilde t^{ab}_0$, 
it suffices to consider the six root generators of $\msu(3)$, which we denote 
by 
\begin{align}
 T_1^+ = \begin{pmatrix}
          0 &\one_2& 0 \\
          0 &0 &0 \\
          0 &0 &0
           \end{pmatrix}, \qquad 
 T_2^+ = \begin{pmatrix}
          0 &0& 0 \\
          0 &0& \one_2 \\
          0 &0& 0
           \end{pmatrix}, \qquad 
 T_3^+ = \begin{pmatrix}
          0 &0 &\one_2 \\
          0 &0 &0 \\
          0 &0 &0
           \end{pmatrix} = [T_1^+,T_2^+]
\end{align}
and similarly $T_i^- = (T_i^+)^\dagger$.
We note that
\begin{equation}
\begin{aligned}
 [T_1^+,\diag(\one_2,-\one_2,\one_2)] &= -2 T_1^+, \qquad  
[T_2^+,\diag(\one_2,-\one_2,\one_2)] = 2 T_2^+,  \\
 [T_3^\pm,\diag(\one_2,-\one_2,\one_2)] &= 0 \ .
\end{aligned}
\end{equation}
This means that $SU(3)$ rotations generated by $\l = z_1 T_1^+ + z_1^* T_1^- + 
z_2 T_2^+ + z_2^* T_2^-$ will 
generate a 4-dimensional orbit through $\tilde t^{ab}_0$ parametrized by 
$(z_1,z_2) \in \C^2$. 
This  provides the missing 4 local coordinates on $\cO_\L \sim \C P^3 \times \C 
P^2 $, complementing the
local $S^4\times S^2$ description of $\C P^3$.
However, this local patch of $\C P^2$ is mapped\footnote{Observe that $t^5 = 0$ 
is not changed by $SU(3)$, which 
stabilizes the base point $x_N$ of $\C P^3$.} by $(t^a)$ degenerately to 
$\R^2_{34} \subset \R^4$ via $t^1 = t^2 = 0$, $t^4+i t^3 = z_2$,
since the $T_1^\pm$ direction are not seen by $t^a$.
These  missing $T_1^\pm$ directions could be resolved by including, for 
example,
\begin{align}
 s^3 = \tilde t^{13}, \quad  s^4 = \tilde t^{14}
\end{align}
as extra embedding coordinates. Then $(t^1,t^2,t^3,t^4,s^3,s^4)$ 
describes precisely the squashed $\C P^2$ as discussed in 
\cite{Steinacker:2015mia,Steinacker:2014lma}, which has a triple 
self-intersection at 
the origin as in Figure \ref{fig:squashed-CP2a}.
Projecting out the $s^{3,4}$ leads to a further projection along 2 of the 6 
directions.

To see the global structure, consider the Weyl group $\cW$ of $SU(3)$, 
which acts by permuting these $2\times 2$ matrix blocks in \eqref{SU(3)-t}. 
$\cW$ is generated by three reflections $w_{1,2,3}$, which
map $\xi$ to 2 other points $w_1\xi, w_3\xi$ as follows
\begin{align}
 w_1 \tilde t^{ab}_0 = \diag(-\one_2,\one_2,\one_2),\qquad w_2 \tilde t^{ab}_0 = 
\tilde t^{ab}_0,  
 \qquad w_3 \tilde t^{ab}_0 = \diag(\one_2,\one_2,-\one_2) \ .
\end{align}
By inspection, $\cW$ preserves $x_N$ and $x_n$ 
at the reference point $\xi$, which means that these 3 distinct points on 
$\cO_\L$ have the same coordinates $(x_N, x_n, t)$.
This shows the 3-fold covering structure near $t=0$. We have already seen that
the local patch  of $\C P^2$ near $\xi$  is mapped by $t$ to $\R^2_{34} \subset 
\R^4$, 
providing a degenerate partial cover of the $t$ coordinates near 0.
Similarly, the local patch near $w_1 \xi$ described by $SU(3)$ rotations 
generated by 
$\l = z_1 T_1^+ + z_1^* T_1^- + z_3 T_3^+ + z_3^* T_3^-$ 
is given by
\begin{equation}
\begin{aligned}
 [T_1^+,\diag(-\one_2,\one_2,\one_2)] &= 2 T_1^+, \qquad  
[T_3^+,\diag(-\one_2,\one_2,\one_2)] = 2 T_3^+,  \\
 [T_2^\pm,\diag(-\one_2,\one_2,\one_2)] &= 0 \ ,
\end{aligned}
\end{equation}
which is mapped by $(t^a)$ degenerately to 
$\R^2_{12} \subset \R^4$ via $t^1 +i t^2 = z_3$, $t^4 = t^3 = 0$. 
In contrast,  the local patch near $w_3 \xi$ described by $SU(3)$ rotations 
generated by 
$\l = z_2 T_2^+ + z_2^* T_2^- + z_3 T_3^+ + z_3^* T_3^-$ is given by
\begin{equation}
\begin{aligned}
 [T_2^+,\diag(\one_2,\one_2,-\one_2)] &= -2 T_2^+, \qquad  
[T_3^+,\diag(\one_2,\one_2,-\one_2)] = -2 T_3^+,  \\
 [T_1^\pm,\diag(\one_2,\one_2,-\one_2)] &= 0 \ ,
\end{aligned}
\end{equation}
which is {\em non-degenerately} mapped  to $\R^4$ by $t^1 +i t^2 = z_3$, $t^4+i 
t^3 = z_2$.

To summarize, the map $t: \cO_\L \to \R^4$ 
near $t=0$ is a triple covering as illustrated in Figure 
\ref{fig:squashed-CP2b}, which is a projection of squashed $\C P^2$  to 
$\R^4$.

This self-intersecting structure is very interesting. 
It suggests that the low-energy physics on  $S^4$ 
will have 3 generations, which arise from fermionic strings connecting these 3 
sheets at the origin $t=0$.
This is very close to the situation studied in 
\cite{Steinacker:2015mia,Steinacker:2014lma},
where a similar squashed fuzzy $\C P^2$ 
led to 3 generations with low-energy physics not far from the Standard Model.
Interestingly, here we would expect 2+1 generations, since the string 
connecting 
the two degenerate sheets is different from the 2 strings connecting one 
degenerate sheet with 
the regular sheet.
As in \cite{Steinacker:2015mia,Steinacker:2014lma}, these fermionic modes are 
expected to be 
chiral\footnote{At least upon switching on additional 2 scalar fields taking 
VEV's 
given by $s^3,s^4$ as above.}.
It is tempting to relate this to the fact that there are 2 light and 1 heavy 
generations in the Standard 
Model.
This is one of the intriguing aspects of the present background, further 
improving the picture of
\cite{Steinacker:2015mia,Steinacker:2014lma}.
\subsection{Poisson brackets}
We define 
\begin{align}
 \tilde x^a = \tilde m^{a6} = (\m^3)^{a6} \ .
\end{align}
From the basic Poisson brackets \eqref{Poisson-m}, we obtain 
\begin{equation}
\label{Poisson-xxt}
\begin{aligned}
\{x^a, \tilde x^c\} &= \tilde m^{ac} \; , \\
\{t^a,x^c\} &= t^{ac}  -\cR^2 g^{ac} = \tilde t^{ac}  - 2 \delta\, g^{ac} \; , 
\\
\{t^a, t^c\} &= -\cR^2 \, m^{ac} -\tilde m^{ac} - x^{a} t^{c} + t^{a} x^c \; , 
\\
  \{t^a, \tilde x^c\} 
  &=  \frac 12(N^2+n^2) \big(\cR^2\, g^{ac} - t^{ac}\big) + \tilde x^{a} x^c  - 
x^{a} \tilde x^{c} \; ,
\end{aligned} 
\end{equation}
for $a,c \neq 6$. Here, $\delta= x_N\cdot x_n$ as defined in 
\eqref{xNn-matrix-class}, and we used 
\begin{align}
 \tilde x \cdot x &= -\frac 12(N^2+n^2)\cR^2 + \frac 1{16}(N^2-n^2)^2 \ .
\end{align}
which follows from \eqref{char-generic-class-2}.

To compare with the fuzzy case, we also compute the following double Poisson 
brackets:
\begin{subequations}
\begin{align}
\sum_{a=1}^5 \{x^a, \{x^a,x^c\}\} =& -4 x^c \; , 
\label{eq:Laplace_eigenvalue_x_classic}\\
 \sum_{a=1}^5 \{t^a,\{t^a,t^c\}\} =& -\left( 3(N^2+n^2) +\Tr(t) \right) t^c + 
12 \cR^2 t^c  \nn \\
=& -\frac{1}{2} \left( 9N^2 +2Nn +9n^2 \right)  t^c +12 \cR^2 t^c  \; .
\label{eq:Laplace_eigenvalue_t_classic}
\end{align}
\end{subequations}

As emphasized before, the Poisson structure lives on the bundle space $\cO_\L$, 
and does not respect the various 
projections  to $S^4$.
However, it can be viewed as a bundle of Poisson structures on $S^4$, giving 
rise to a frame bundle on $S^4$ as 
emphasized in \cite{Steinacker:2016vgf}.
\paragraph{Radial constraint and $y$ variables.}
We can now identify generators which commute with $\cR^2$. 
This would allow to impose the radial constraint 
\begin{align}
 \cR^2 = \mathrm{const} \,,
\end{align}
which defines a generalized irreducible $SO(5)$ sphere.
From the above Poisson brackets, we obtain
\begin{equation}
 \label{R2-Commutators}
\begin{aligned}
 \{\cR^2, x^a\} &= 2 x_b m^{ba} = 2 t^a  \; ,\\
 \{\cR^2, \tilde x^a\} &= 2 x_b \tilde m^{ba} = - (N^2+n^2) t^a \; ,\\
 \{\cR^2, t^a\} &= 2 x_b (t^{ba} -\cR^2 g^{ba}) = -2 \tilde x^a - 2\cR^2 x^a \; 
, \qquad a \neq 6
\end{aligned}
\end{equation}
and we see that the following generators $y^a$ commute with $\cR^2$:
\begin{equation}
 \label{y-def}
\begin{aligned}
 \{\cR^2, y^a\} &= 0 \ ,  \\
 y^a &\coloneqq \frac 2{N^2-n^2}\big((N^2+n^2) x^a + 2 \tilde x^a\big) = x_N^a 
- x_n^a \ .
\end{aligned} 
\end{equation}
This means that wave functions $\phi(y^a)$ strictly respect $\cR^2$. 
In particular, a semi-classical Laplacian chosen as 
\begin{align}
 \Box_y \coloneqq -\{y^a,\{y_a,\cdot\}\}
\end{align}
(compare to fuzzy case \eqref{eq:def_Laplacian}) induces no kinetic term 
in the radial dimensions, and  $\Box_y y^a$ Poisson-commutes with $\cR^2$. 
Therefore,
\begin{align}
 \Box_y y^a = c_1 y^a + c_2 \cR^2 y^a \ 
 \label{Box-y-class}
\end{align}
for suitable constants $c_1, c_2$. This corresponds to  
\eqref{eqn:sol_spherical_XN-Xn} in 
the fuzzy case.
Furthermore, the above Poisson brackets imply
\begin{subequations}
\begin{align}
 \{t^a,y^b\} 
   &= y^a x^b - x^a y^b \; ,  \label{ty-yb}  \\
\{y^a,y^b\} 
&=  \frac{4}{N^2-n^2}\,\left(\left(\frac{N^2+n^2}{2} - \cR^2\right) y^{ab}
   + y^a t^b - t^a y^b\right) \; ,
\end{align}
where 
\begin{align}
 y^{ab} \coloneqq m_N^{ab}-m_n^{ab}
\end{align}
\end{subequations}
in analogy to \eqref{y-def}.
In particular, this means that 
\begin{align}
 \{t^\mu,y^\nu\} = 0, \qquad \mu,\nu = 1,\ldots,4
 \label{tmu-ynu}
\end{align}
in tangential coordinates where e.g. $x^a \sim e_5$.
Thus, $t$ can be viewed as momentum generator for either of $x_N$, $x_n$ or 
$x$, but not for $y$.

We also note the following constraints:
\begin{equation}
\begin{aligned}
 \{y^a,y^b\} y_b &= 0 \ ,  \\
  m^{ab} y_b &= m^{ab} (x_b + \frac{2}{N^2+n^2} \tilde x^a_b) = 0
\end{aligned}
\end{equation}
from \eqref{char-generic-class-2}. 
This means that imposing $\cR^2 = \mathrm{const}$ is incompatible 
with the gravity mechanism discussed in \cite{Steinacker:2016vgf}, as for the 
basic fuzzy 4-sphere. 
We will, therefore, not impose this constraint.
In contrast, $t^a$ does not commute with $\cR^2$.

There is an interesting alternative approach to the $y^a$ generators:
Consider
\begin{align}
 e^{ab} = \epsilon^{abcdef}m^{cd}m^{ef}, \qquad    e^a = e^{a6} \; .
\end{align}
By anti-symmetry, $e^{ab}$ must be a linear combination of $m^{ab}$ and $\tilde 
m^{ab}$; hence, 
$e^{a}$ must be a linear combination of $x_N^a$ and $x_n^a$.
Since it Poisson-commutes with $\cR^2$, it follows that
\begin{align}
 y^a \sim e^a \ .
\end{align}

\subsection{Summary of Poisson brackets}
To facilitate the comparison with the fuzzy case, we collect the basic 
relations for the 
Poisson brackets on the classical orbit $\cO_\L$ here:
\begin{equation}
 \label{Poisson-brackets-collect}
\begin{aligned}
 \{x^a, x^b\} &= m^{ab} \; , \\
 \{x^a, y^b\} &= y^{ab} \coloneqq m_N^{ab} - m_n^{ab} \; , \\ 
 \{x^a,t^b\} &= -\tilde t^{ab}  + 2 \delta\, g^{ab} \; ,  \\
 \{y^a,y^b\} &=  \frac{4}{N^2-n^2} \,\left(\left(\frac{N^2+n^2}{2} - 
\cR^2\right)  y^{ab} + y^a t^b - t^a y^b \right) \; , \\
 \{t^a, t^b\}
  &=  \Big(\frac{N^2+n^2}{2} - \cR^2 \Big)\, m^{ab}  -\frac{N^2-n^2}{4}\, 
y^{ab} 
- x^{a} t^{b} + t^{a} x^b \; , \\
 \{t^a,y^b\} &= y^a x^b - x^a y^b\; , 
\end{aligned}
\end{equation}
where $a,b = 1,\ldots,5$. 
%
%
\subsection{Functions on \texorpdfstring{$\cS^4_\L$}{S4}}
\label{sec:funct-on-S4}
\paragraph{Local coordinates.}
We restrict ourselves to  generic points where $x_N$ and $x_n$ are linearly 
independent.
As discussed in Section \ref{sec:global}, this means that  $x_N$, $x_n$, and $t$
provide a good local description of $\cO_\L$.
Hence, all functions can be represented locally as $\phi(x_N,x_n,t)$.
In particular, the $5\times 5$ matrices $m_N^{ab}, \tilde m_n^{ab}$ and $t^{ab}$ 
can be expressed in terms of  $x_N,x_n,t$ as follows: 
Their action on $x_N,x_n,t$ is obtained from \eqref{MNnO-vector-algebra-class}, 
which implies that  $m_N$ and $m_n$
respect this $3\times 3$  block-structure.
The remaining antisymmetric $2\times 2$ block can then be written in terms of 
\begin{align}
 f^{ab} &:= \epsilon^{abcde} x_N^{c} x_n^{d} t^{e} \ = - f^{ba} \ ,  \quad 
\qquad
 0 = f_{ab} x_N^a = f_{ab} x_n^a = f_{ab} t^a  \ .
\end{align}
Explicitly, one finds 
\begin{subequations}
\begin{align}
 m_{N}^{ab} &= \frac{2}{t\cdot t} (v_N^a t^b - t^a v_N^b) + d_N f^{ab} \,, \\
 m_{n}^{ab} &= \frac{2}{t\cdot t} (v_n^a t^b - t^a v_n^b) + d_n f^{ab} \,,
\end{align}
where 
\begin{align}
 v_N &= \frac{N^2}{4} x_n - \d\ x_N \,, \qquad v_N \cdot x_N = 0 \,, \quad v_N 
\cdot v_N = \frac{N^2}{16} t\cdot t \,,  \\
 v_n &= \frac{n^2}{4} x_N - \d\ x_n \,, \ \qquad v_n \cdot x_n \ = 0 \,, \ 
\quad v_n \cdot v_n \ = \frac{n^2}{16} t\cdot t  \,,
\end{align}
\end{subequations}
and $d_{N,n}$ can be computed from the (anti)self-duality of $m_{N,n}$ at 
$x_{N,n}$. 
The matrix $t^{ab}$ could be determined similarly.
However, this description does not work if $x$ and $y$ are parallel.
We can then use, for instance, $m^{ab}$ to complement $x$ and $t$, which may be 
useful also more generally.

Explicit local coordinates on  $\cO_\L$ can be found as follows:
Since $\cO_\L$ is a homogeneous space, we can pick any given reference point $p 
\in \cO_\L$.
Such a reference point could be\footnote{We drop the superscripts in 
$R_N^{(c)}$ 
because we are in the semi-classical regime anyway.} 
\begin{align}
 x_N = e_5 R_N, \quad  x_n = R_n(\cos\vartheta e_5 + \sin\vartheta e_4), \quad \ 
t = R_t e_3
 \label{ref-point-Otilde}
\end{align}
where $R_t$ can be extracted from \eqref{constraints-explicit}.
We can then use $x_N^{1,\ldots,4}$, $x_n^{1,2,3}$, and $t^{1,2}$ as local 
coordinates. 
The missing $10^{\mathrm{th}}$ coordinate is provided by $\vartheta$.

Finally, we note that  the only $SO(5)$-invariant functions on $\cO_\L$
are given\footnote{A priori, there are the two $\mso(5)$ 
Casimirs, $C_2[\mso(5)] \sim  C_2[\mso(6)] - 2\cR^2$ and 
$C_4[\mso(5)] \sim e\cdot e$, where $e^a = \varepsilon^{abcde6}m^{bc}m^{de} = 
e^{a6}$.
By anti-symmetry, $\cE^{a6}$ must be a linear combination of $m^{a6}$ and 
$\tilde m^{a6}$, hence 
$\cE^{a}$ must be a linear combination of $x_N^a$ and $x_n^a$.
Since it Poisson-commutes with $\cR^2$, it must be proportional to $y^a$,
and $y\cdot y$ reduces to $\cR^2$.
The same argument applies in the fuzzy case, where $\cR^2$ generates all 
$SO(5)$-invariant
operators in $\End(\cH)$.}  by $f(\cR^2)$.
\paragraph{Bundle structure and higher spin.}
We want to describe the fluctuation modes on $\cO_\L$ in a $SO(5)$-covariant 
way, reflecting 
the local bundle structure  $S^4 \times \cK$. 
We will consider $S^4 \ni x$ 
as physical configuration space, and use $\xi$ to describe points on $\cK$ in 
the following.

In standard Kaluza-Klein (KK) compactification, the harmonics on $\cK$ yield a 
KK tower of {\em scalar} fields on $S^4$. 
However, in the present situation $SO(5)$ acts  on both $S^4$ and $\cK$, which 
form a twisted (equivariant) bundle.
This means that functions $\phi(x,\xi) \in \cC(S^4\times \cK) \cong 
\cC(S^4)\otimes \cC(\cK)$ 
must be decomposed as tensor product of $SO(5)$ reps, 
which leads to a tower of {\em higher spin modes} on $S^4$ instead of the usual 
KK tower.
This twisted structure  is the crucial point of these covariant backgrounds.
It transmutes KK modes into, for example, gravitons and other higher spin 
modes with the appropriate local transformation properties.
\paragraph{Mode expansion.}
With this in mind, we want to represent functions on $\cO_\L$ as functions on 
$S^4$ expanded into the harmonics on $\cK$.
An $SO(5)$-covariant way to organize functions on $S^4\times \cK$ is as 
follows\footnote{As discussed above,
we can trade $y^a$ for either $m^{ab}$ or $\tilde m^{ab}$.}:
\begin{align}
 \phi &=  \phi(x,t,y)  \nn\\
 &= \phi(x,t) + \phi_{ab}(x,t)m^{ab} + \phi_{ab;cd}(x,t)m^{ab} m^{cd}+ \ldots 
\; ,
\end{align}
see for instance \cite{Steinacker:2016vgf}.
These modes can be expanded further in polynomials in $t$ as 
\begin{align}
 \phi_{ab}(x,t) =  \varphi_{ab}(x) +  \varphi_{ab;e}(x) t^e +  
\varphi_{ab;ef}(x) t^e t^f + \ldots
 \label{phi-t-expand}
\end{align}
etc. The  $\varphi_{ab;e}(x)$ etc. clearly describe some higher spin theory on 
$S^4$, and 
it would be desirable to make contact with (Vasiliev-type \cite{Vasiliev:2004qz}) 
higher spin theories\footnote{This  even applies for non-compact homogeneous spaces in a Lorentzian setting, 
which
might resolve the problem of averaging over such an internal space  encountered 
in \cite{Doplicher:1994tu}.}.
To understand the role of possible radial contributions, consider e.g.
 $\phi_{ab}(x) = h_a(x) x_b - h_b(x) x_a$. Then 
\begin{align}
 \phi_{ab} m^{ab}  &= (h_a(x) x_b - h_b(x) x_a) m^{ab}  = -2 h_a t^a 
\end{align}
recalling $t = -\m \cdot x$. Hence, radial components of $\phi_{ab}$ give rise 
to $t^a$, 
which is included in \eqref{phi-t-expand}.
To avoid over-counting, $\phi_{ab;cd}$ should, therefore, be tangential on 
$S^4$ and traceless.
\paragraph{Dimensional reduction.}
The above $SO(5)$ organization is not sufficient to guarantee a truly 
4-dimensional theory.
The  dynamics is governed by $\Box$ \eqref{eq:def_Laplacian}, which is some
Laplace operator on the 10-dimensional space $\cO_\L$.
An effectively 4-dimensional theory is obtained via dimensional reduction on 
$\cK$:
if only the trivial mode on $\cK$ participates (or dominates), 
then the wavefunctions  are given by $\phi(x)$, depending only on $x$, 
and $\Box$ reduces to a Laplace operator on the large $S^4$.
More precisely, initial configurations which are (almost) constant on $\cK$ 
should remain so
under the dynamics.

There are different possibilities to justify this scenario. One possibility 
is a large mass gap on $\cK$, so that excitations along $\cK$ do not play an 
important role at low energies.
This requires a large asymmetry in $\Box$. While this is natural for
$n \ll N$,  it  depends on the  embedding of the background in the 
matrix model, which can be reliably addressed only once
the effective potential including quantum corrections is sufficiently well 
understood.

We will encounter also another -- in a sense dual -- mechanism for dimensional 
reduction:
Assume that only 5 matrices are embedded, as in the background $ \cY^A \sim 
( T^a , 0 )^T$
considered in Section \ref{sec:phase-space-embed}. 
Then the effective metric on $\cO_\L$ has reduced rank (rather than full rank 
10)
and is restricted to $x$ space, while 
{\em no} kinetic term in the $y$ direction is induced. The internal modes 
in $y$ space are clearly higher spin 
modes whose propagation is only 4-dimensional.
More generally, embeddings such as $ \cY^A \sim  ( T^a , Y^a )^T$ 
where one scale  is much larger than the other
 would naturally lead to dimensional reduction.

Dimensional reduction is further supported by the fact that excitations on $\cK$ 
couple only via derivatives
to functions on $S^4$, which is suppressed at low energies by the scale of 
non-commutativity. These non-trivial $\cK$ modes 
include the spin 2 gravitons on $S^4$, and the higher spin modes will be suppressed even more.
However this is a complicated issue which can be settled only 
if the background geometry is known.
\paragraph{Averaging.}
To compute the low-energy observables in this dimensional reduction scheme, one 
needs to project on
functions which are constant in $\cK$. This can be done by ''averaging`` 
over $\cK$ as follows: Consider any given point (the ''north 
pole``) $p \in S^4$. 
As discussed before, the fiber $\cK$ over $p$
is the orbit of its stabilizer $SO(4) \subset SO(5)$. Thus, averaging over this 
fiber is tantamount to averaging over the 
local $SO(4)$ stabilizer, and is denoted by
\begin{align}
 [f(x,\xi)]_0 \ .
\end{align}
By construction, the result is a scalar of the local stabilizer $SO(4)$.
Carrying this out over each point, one obtains e.g.
\begin{align}
 [t^{ab}]_0 = \frac 14 N^2 \left(g^{ab} + \cO\left(\frac{1}{N}\right)\right)\ , 
\qquad  [m^{ab}]_0 = 0 = [y^{ab}]_0 \ ,
\qquad a,b = 1,\ldots,5 \; .
\end{align}
%
%
%
%
\section{Fuzzy geometry}
\label{sec:fuzzy_geometry}
The results of the previous sections provide the guideline for 
organizing the corresponding fuzzy algebra, and for finding embeddings which are 
solutions 
of IKKT-type matrix models. 
\subsection{Fuzzy operator algebra}
The algebra $\End(\cH)$ of functions on fuzzy $\cS^4_\L$ is properly 
understood as quantized algebra 
of functions on  $\cO_\L$, which decomposes into the direct sum of higher spin 
harmonics on $S^4$.
For the basic fuzzy sphere $S^4_N$ with $\L =(N,0,0)$, the underlying orbit is 
$\C P^3$, and the fuzzy algebra of functions is
\begin{align}
 \End(\cH_N) = (0,0,N)\otimes (N,0,0)  
 &=  \bigoplus\limits_{n\leq N} (n,0,n)_{\mso(6)} \nn\\
 &\cong \bigoplus\limits_{m\leq n\leq N} (n-m,2m)_{\mso(5)}  \ ,
 \label{mode-expansion-fuzzy-alg}
\end{align}
cf. \cite{Steinacker:2015dra}.
This is the twisted analog of a KK tower with intrinsic UV cutoff; 
for example, $\cM^{ab} \in (1,0,1)$.
For the generalize fuzzy spheres with $\L = (N,0,n)$, additional modes arise, 
and
some multiplicities become non-trivial. It turns out that the 
following structure holds  for the lowest modes:
\begin{align}
 \End(\cH_\L) &= (n,0,N)\otimes (N,0,n) 
 = (0,0,0) \oplus 2(1,0,1) \oplus (0,2,0) \oplus \ldots 
  \label{mode-expansion-general}
\end{align}
provided $N,n > 0$. 
\paragraph{Automorphism.}
There is an involutive anti-linear automorphism given by 
\begin{subequations}
\begin{equation}
 \sigma(\cM^{ab}) = - \cM^{ab} \, , \qquad \sigma(i) = -i \, ,
\end{equation}
which satisfies 
\begin{equation}
  \sigma(\tilde\cM^{ab}) = - \tilde\cM^{ab}\,  , \qquad \sigma(\cT^{ab}) = 
\cT^{ab} \, .
\end{equation}
\end{subequations}
On $\End(\cH)$, this automorphism is nothing but complex conjugation, but 
it helps to understand better the  algebraic properties in the fuzzy case,
such as $\Box T \sim T$ etc.
%
%
\paragraph{Matrix operators.}
We will now identify the fuzzy analogs $\cM^{ab}$, $\tilde \cM^{ab}$,
and $\cT^{ab}$ of the anti-symmetric resp. symmetric matrix-valued functions 
$m^{ab}$, $\tilde m^{ab}$,
and $T^{ab}$. The operators are defined by the same (anti)symmetry and trace 
conditions as their classical counterparts.
First, $\cT^{ab}$ is defined to be the symmetric traceless part of $\cM^2$,
\begin{align}
 \cT^{ab} \coloneqq -(\cM^2)^{\{ab\}} 
  &= \frac 12 \{\cM^{ac},\cM^{bc'}\}_+ \ g_{cc'}  = -(\cM^2)^{ab} -2 i\cM^{ab} 
\, ,
\end{align}
using the Lie algebra relations in the last line.
It then follows that
\begin{subequations}
\begin{align}
 (\cT^{ab})^* = \cT^{ab} = \cT^{ba}
\end{align}
and 
\begin{align}
 \cT^{66} = \cR^2 \,.
\end{align}
\end{subequations}
$\cT^{ab}$ is, in fact, the unique $(0,2,0)$ tensor operator in
$\End(\cH_\L)$ for $\L = (N,0,n)$,
and it vanishes for $\L = (N,0,0)$.

Similarly, $\tilde\cM^{ab}$ is defined to be the anti-symmetric part of $\cM^3$,
\begin{equation}
\begin{aligned}
 \tilde\cM^{ab} \coloneqq (\cM^3)^{[ab]} &= \frac 12( 
\cM^{ac}\cM^{cd}\cM^{db} - \cM^{bc}\cM^{cd}\cM^{da}) \\
  &=  (\cM\cM\cM)^{ab} + 5 i (\cM\cM)^{ab} - 10 \cM^{ab} -\frac i2  
(\Tr\cM^2)\, 
g^{ab} \, ,
\end{aligned}
\end{equation}

using again the Lie algebra relations in the last line.
Again, it follows that
\begin{align}
 (\tilde\cM^{ab})^* = \tilde\cM^{ab} = -\tilde\cM^{ba} .
\end{align}
The story of independent monomials of $\cM$ ends here, since $\cM^4$ can be 
expressed 
in terms of the above matrix generators 
via its characteristic equation. This arises as follows:
\paragraph{Characteristic equation.}
Since $\End(\cH_\L)$ is a quantization of the algebra of functions on $\cO_\L$,
its decomposition into harmonics is the same below the cutoff $N$. For the 
generalized spheres, we note that 
the decomposition \eqref{mode-expansion-general}
contains only two $(1,0,1)$ operators. These must be given by $\cM^{ab}$ and 
$\tilde\cM^{ab}$.
Furthermore, there is only one $(0,2,0)$ operators, which must be $\cT^{ab}$.
Therefore, $\cM$, $\cT$, $\tilde\cM$, and $g$ form a closed matrix algebra. 
In other words, $\cM$ satisfies 
a characteristic equation of order 4, which is the matrix analog of 
\eqref{char-generic-class}.
For $\L=(N,0,n)$, it takes the form (see Appendix \ref{app:chareq} for a 
derivation)
\begin{align}
 \left(\left(i\cM+2\right)^2 - \frac{(N-n)^2}{4}\right)
 \left(\left(i\cM+\frac{3}{2}\right)^2 - \frac{(N+n+3)^2}{4}\right) = 0 \ .
 \label{char-eq-fuzzy}
\end{align}
This is clearly consistent with \eqref{char-generic-class} up to quantum 
corrections.
Accordingly, there are two independent Casimirs, given by 
\begin{align}
 C_2[\mso(6)] = -\frac {1}{2} \Tr(\cM^2) = \frac{1}{2} \Tr(\cT), \qquad 
C_3[\mso(6)] = \Tr(\cM^3) ,
\end{align}
while the higher Casimirs reduce to the above via the characteristic equation.
Moreover, the characteristic equation implies
\begin{align}
 \tilde\cM\cM = \cM \tilde\cM = -2i \tilde\cM + R_{\tilde \cM \cM}^T \cT + 
R_{\tilde \cM \cM}^g\, g \ .
 \label{M-tilde-M} 
\end{align}
The coefficients of the anti-symmetric terms $\cM$, $\tilde \cM$ on the rhs  
follow directly from
\begin{align}
 \cM_{ac} \tilde\cM_{cb} - \tilde\cM_{bc} \cM_{ca} &= [\cM_{ac}, \tilde\cM_{cb}]
   = -4i \tilde\cM^{ab} \,,
\end{align}
which implies that there is no $\cM$ term in \eqref{M-tilde-M}.
The remaining coefficients $R_{\tilde \cM \cM}^T, R_{\tilde \cM \cM}^g$ are 
found to be
\begin{subequations}
\begin{align}
 R_{\tilde \cM \cM}^T &= -\frac{1}{2} \left(N^2 +n^2 +3N +3n +8\right) , \\
 R_{\tilde \cM \cM}^g &=-\frac{1}{16} (N-n)^2 (N+n+4) (N+n+2) \nn\\
   &= \ \frac 16\, \left(\Tr\left(\cM^4\right)+ \left(R_{\tilde \cM \cM}^T 
-10\right)\, \Tr(\cM^2)\right) \; .
\end{align}
\end{subequations}
Now, one can work out the full matrix multiplication algebra of the 
$\tilde\cM$, $\cT$, $\cM$, and $g$.
This will be given for the modified basis $\cM_N$, $\cM_n$ as determined below. 
\paragraph{Vector operators.}
As in the commutative case, we can define the $SO(5)$ vector operators 
\begin{align}
 X^a &= \cM^{a6}, \qquad \tilde X^a = \tilde\cM^{a6}, \qquad  T^a = 
\tilde\cT^{a6} \ 
\end{align}
for $a=1,\ldots,5$,
which provide possible embeddings of generalized fuzzy spheres in the matrix 
model.
We can compute their scalar products by taking the $66$ component of 
the above matrix multiplication algebra. 
For example, \eqref{M-tilde-M} implies
\begin{align}
 \tilde X \cdot X = -R_{\tilde \cM \cM}^T \, \cR^2 - R_{\tilde \cM \cM}^g =   X 
\cdot \tilde X \,.
\end{align}
Similarly, one finds 
\begin{align}
X\cdot T + T \cdot X &= 0 = \tilde X\cdot T + T \cdot \tilde X \,,
\label{X-Y-T-orthogonality}
\end{align}
which is consistent with the semi-classical limit $x\cdot t = 0$, see 
\eqref{scalar-products-class}.
The last relations state that $T^a$ is orthogonal to both $X^a$ and $\tilde 
X^a$.
This follows immediately from
\begin{align}
 X\cdot T  &= -\cM^{6a} \cT^{a6}  = -(\cM \cT)^{66}
  = -(\cT\cM)^{66} = -T\cdot X , 
\end{align}
and similarly for $\tilde X$.
\paragraph{Matrix multiplication algebra.}
As in the classical case \eqref{MNn-trafo-class}, we would like to find 
operator-valued $6\times 6$ matrices  
\begin{align}
 \cM_N^{ab}, \qquad \cM_n^{ab}, \qquad \tilde \cT^{ab} \coloneqq \cT^{ab} - c 
g^{ab}\,, 
 \end{align}
where
 \begin{align}
 \begin{pmatrix}
  \cM_N \\ \cM_n 
 \end{pmatrix}
&\coloneqq
\begin{pmatrix} 
 \alpha_N & \beta_N \\  \alpha_n & \beta_n 
\end{pmatrix}
 \begin{pmatrix}
  \tilde{\cM} \\ \cM 
 \end{pmatrix}
 \equiv
 A
  \begin{pmatrix}
  \tilde{\cM} \\ \cM 
 \end{pmatrix} 
 \; .
\label{N-Matrix-def}
\end{align}
The associated vector operators 
\begin{align}
 X_{N}^a &= \cM_{N}^{a6}\, , \qquad \tilde X^a_n = \tilde\cM^{a6}_n \, , \qquad 
\tilde  T^a = \tilde\cT^{a6} = T^a \, 
\label{X-Matrix-def}
\end{align}
for $a=1,\ldots,5$ are defined such that the inner products of the $X_{N,n}$ take the 
simple form\footnote{To see that this is possible, it suffices 
to verify that the coefficient matrix of $\cR^2$ has signature $(1,-1)$, to 
bring it to the form
 $\sim \begin{pmatrix}
        0 & 1\\
        1 & 0
       \end{pmatrix}$, and to kill the off-diagonal real entries by subtracting 
the appropriate $c$.} 
\begin{align}
 \begin{pmatrix}
  X_N\cdot X_N & X_N\cdot X_n \\
   X_N\cdot X_n & X_n\cdot X_n  
 \end{pmatrix}
= 
\begin{pmatrix}
 R_N^2 & 0 \\
 0 & R_n^2
\end{pmatrix}
 + \Delta \begin{pmatrix}
        0 & 1\\
        1 & 0
       \end{pmatrix}, \qquad \Delta = \frac{1}{2}(\cR^2 - c\one)
 \label{inner-products-Nn}
\end{align}
for some suitable $c\in\R$. This is the fuzzy analog of 
\eqref{xNn-matrix-class}, and
it means that $X_{N}$ and $X_n$ describe two  fuzzy 4-spheres with sharp radii 
$R^2_{N}$ and $R^2_n$,
respectively. 

The coefficients $\a_{N,n}$, $\b_{N,n}$ are to be defined such that 
\eqref{inner-products-Nn} holds. 
To carry this out, it is convenient to first rewrite the matrix algebra  of 
$\cM$, $\tilde\cM$, and $\tilde\cT$
in terms of the new operators  $\cM_N$, $\cM_n$, and $\tilde\cT$. Then the 
requirement \eqref{inner-products-Nn} is tantamount to the vanishing of 
the three red coefficients in the following multiplication table\footnote{The 
vanishing of the blue coefficients is a non-trivial result 
of the full calculation \eqref{Rxyz-full}.}:
\begin{equation}
\label{MNnO-algebra}
\begin{aligned}
 \cM_N\cM_N &= i R_{NN}^N\cM_N + i R_{NN}^n\cM_n   {\color{red}{\; +0 \cdot 
\tilde\cT\;}} + R_{NN}^g g \; ,\\
 \cM_n\cM_n &= i R_{nn}^N\cM_N + i R_{nn}^n\cM_n   {\color{red}{\; +0 
\cdot \tilde\cT\;}} + R_{nn}^g g \; , \\
  \cM_N\cM_n = \cM_n\cM_N &= i R_{Nn}^N \cM_N + i R_{Nn}^n \cM_n  + R_{Nn}^T 
\tilde\cT  {\color{red}{\; +0 \cdot g\;}} \; , \\
 \tilde\cT\cM_N &=  {\color{blue}{\; +0 \cdot \cM_N\;}} + R_{TN}^n\cM_n + i 
R_{TN}^T\tilde\cT 
+  i R_{TN}^g\ g  \; ,
\\
 \tilde\cT\cM_n &= R_{Tn}^N\cM_N  {\color{blue}{\; +0 \cdot \cM_n\;}} + i 
R_{Tn}^T\tilde\cT 
+ 
i R_{Tn}^g \ g \; ,
 \\
 \tilde\cT\tilde\cT &= i R_{TT}^N\cM_N + i R_{TT}^n\cM_n +  R_{TT}^T\tilde\cT + 
R_{TT}^g \ g \; .
\end{aligned}
\end{equation}
Requiring that $R_{NN}^T = 0 = R_{nn}^T$ yields the solutions 
\begin{equation}
\begin{aligned}
 \beta_N &= \frac{\alpha_N}{4}  \left(
 2 N^2+6 N + 2 n^2 +6 n+16
 +\kappa  (N-n) 
\sqrt{(N+n+4)(N+n+2)}\right)  \; ,\\
 \beta_n &= \frac{\alpha_n}{4}  \left(
 2 N^2+6 N + 2 n^2 +6 n+16
 +\gamma  (N-n) 
\sqrt{(N+n+4)(N+n+2)}\right) \; , 
\end{aligned}
\end{equation}
for $\kappa, \gamma = \pm1$.
For $A$ to be invertible, it follows from
\begin{subequations}
\begin{align}
 \det{A} 
 &= \frac{1}{4} \alpha_n \alpha_N (\gamma -\kappa ) 
(N-n) \sqrt{(N+n+4)(N+n+2)} \;   
\end{align}
that one has to choose  $\gamma \neq \kappa$. Thus, $\gamma = - \kappa$ and the 
remaining freedom $\kappa = \pm1$  simply 
interchanges $N \leftrightarrow n$. The 
determinant reduces to
\begin{align}
 \det{A}&= \alpha_n \alpha_N \kappa \det{\tilde{A}} \\
\det{\tilde{A}}&= \frac{n-N}{2}\sqrt{(N+n+4) (N+n+2)}  
 \ = \ - 2 \sqrt{-R_{\tilde \cM \cM}^g}\; 
\end{align}
\end{subequations}
where $R_{\tilde \cM \cM}^g$ was defined in \eqref{M-tilde-M}.
Moreover, we observe
\begin{align}
 \frac{\beta_N}{\a_N}  &\to \frac{3 N^2}{4}, \qquad \frac{\b_n}{\a_n} \to 
-\frac{N^2}{4}, \qquad \k=1, N \gg n \, ,
 \label{MNn-trafo-semiclass}
\end{align}
which is consistent with the classical limit \eqref{MNn-trafo-class}. 
Up to now, the $\a_N$, $\a_n$ are arbitrary normalization constants.
To stress the similarity to the classical set-up
\eqref{eq:XN+Xn_classical}, we determine $\a_N$, $\a_n$ by imposing
%
\begin{align}
  X_N + X_n &= X \ ,
\end{align}
which holds for
\begin{align}
 \alpha_N = -\frac{\kappa}{\det{\tilde{A}}} = - \a_n \; , \quad \kappa 
=\pm1 \; .
  \label{eq:fix_alphaN}
\end{align}
Moreover, we determine $c$ in the definition of $\tilde\cT$ via the condition 
$R_{Nn}^g=0$, which  yields
\begin{equation}
 c=\frac{1}{4} \left(N^2 +n^2 +3 N +3 n\right) \; .
\end{equation}
It is useful to note that
\begin{align}
 C_2[\Lambda]-3c&= \frac{1}{4}\left(3N+3n+2Nn \right) \; .
\end{align}
Then the fuzzy transformation matrix \eqref{N-Matrix-def} reads explicitly  
\begin{align}
 A
&= -\frac{\kappa}{\det{\tilde{A}}} 
\begin{pmatrix} 1 & 2(c+2) - \frac{\kappa}{2} \det{\tilde{A}}   \\ 
-1 & -2(c+2) - \frac{\kappa}{2} \det{\tilde{A}} \end{pmatrix} 
= \frac{-1}{\det{\tilde{A}}} 
\begin{pmatrix} 1 & 2(c+2) - \frac{ \det{\tilde{A}}}{2}   \\ 
-1 & -2(c+2) - \frac{\det{\tilde{A}}}{2}  \end{pmatrix} \ .
\end{align}
Here, we have chosen $\kappa =1$, because 
$\det{\tilde{A}}$ is negative for $N> n$.
Then
\begin{align}
\begin{matrix}
\a_N = \frac{2}{(N-n) \sqrt{(N+n+4)(N+n+2)} } &\approx \frac{2}{(N^2-n^2)} 
\\
\a_n = \frac{-2}{(N-n) \sqrt{(N+n+4)(N+n+2)} } &\approx  \frac{-2}{(N^2-n^2)} \\
 \hfill \det A &\approx \frac{2}{N^2-n^2}
\end{matrix}\; ,
\qquad \text{for } N\gg n \gg 1 \; ,
\end{align}
which is consistent with the classical limit \eqref{MNn-trafo-class}.

Having solved the constraints imposed on the algebra \eqref{MNnO-algebra}, one 
can readily evaluate the remaining structure constants. The result is given in 
Appendix \ref{sec:stucture_constants}. 
In particular, we find that the two blue terms in \eqref{MNnO-algebra} vanish 
exactly,
as in the commutative case \eqref{MNnO-vector-algebra-class}.
\paragraph{Commutators.}
To find spherical embeddings of these fuzzy spaces in
matrix models, we will use the  $SO(5)$-vector 
operators \eqref{X-Matrix-def} obtained from $\cM_N, \cM_n,  \tilde\cT $.
The resulting commutator relations are as follows:
\begin{subequations}
\label{eq:commutators_XNXn}
\begin{align}
 \left[X_N^a,X_n^b\right] = \left[X_n^a,X_N^b\right] =& 
\frac{i }{4  \det{\tilde{A}}} 
\left( 4c + \det{\tilde{A}}\right)
M_N^{ab} 
-\frac{i }{2 \det{\tilde{A}}}  M_N^{ab} \cR^2
\label{XNXn-CR} \\*
&-\frac{i  }{4 \det{\tilde{A}} } 
\left(   4c -  \det{\tilde{A}}\right)
M_n^{ab} 
+\frac{i }{2 \det{\tilde{A}}} M_n^{ab} \cR^2 \nn \\*
&+\frac{i }{2 \det{\tilde{A}}}
\left( X_N^a T^b - T^a X_N^b \right) 
-\frac{i }{2 \det{\tilde{A}}}
\left( X_n^a T^b - T^a X_n^b \right) \nn \\*
&+
\frac{1}{\det{\tilde{A}}  }
\left( X_N^a X_n^b - X_n^a X_N^b \right) \nn \; , \\
 \left[X_N^a,X_N^b\right] =& 
 -\frac{i }{4 \det{\tilde{A}} }
 \left( 4  c-3  \det{\tilde{A}}\right)
M_N^{ab} 
+\frac{i }{2 \det{\tilde{A}}}
M_N^{ab} \cR^2  \\*
&+  \frac{i }{4 \det{\tilde{A}} }
\left(4   c- \det{\tilde{A}}\right)
M_n^{ab} 
 -\frac{i }{2 \det{\tilde{A}}} 
M_n^{ab} \cR^2 \nn \\*
&-\frac{i }{2 \det{\tilde{A}}}
\left( X_N^a T^b - T^a X_N^b \right)  
+ \frac{i }{2 \det{\tilde{A}}}
\left( X_n^a T^b - T^a X_n^b \right) \nn \\*
&- \frac{1}{ \det{\tilde{A}}  } 
\left( X_N^a X_n^b - X_n^a X_N^b \right) \; , \nn \\
 \left[X_n^a,X_n^b\right] =& 
 -\frac{i }{4 \det{\tilde{A}} }
 \left(4 c + \det{\tilde{A}}\right)
 M_N^{ab} 
+\frac{i }{2 \det{\tilde{A}} } M_N^{ab} \cR^2 \\ *
&+ \frac{i }{4  \det{\tilde{A}}} 
\left(4c+  3 \det{\tilde{A}}\right)
M_n^{ab} 
-\frac{i }{2 \det{\tilde{A}}} M_n^{ab} \cR^2 \nn \\*
&-\frac{i }{2 \det{\tilde{A}} }
\left( X_N^a T^b - T^a X_N^b \right) 
+\frac{i }{2 \det{\tilde{A}}}
\left( X_n^a T^b - T^a X_n^b \right) \nn \\*
&-\frac{1}{ \det{\tilde{A}} } 
\left( X_N^a X_n^b - X_n^a X_N^b \right) \; , \nn \\
\left[T^a,T^b\right] =&
 + \frac{i }{2 } (\det{\tilde{A}}+4 c ) M_N^{ab}
-  i  M_N^{ab} \cR^2
  \\*
&
-\frac{i }{2 } (\det{\tilde{A}}-4 c ) M_n^{ab}
- i  M_n^{ab} \cR^2
\nn \\*
&+ \tilde\cT^{ab} +cg^{ab} - g^{ab} \cR^2    \nn \\*
& -i \left( X_N^a T^b - T^a X_N^b \right) 
-i \left( X_n^a T^b - T^a X_n^b \right) \; , \nn \\
\left[T^a,X_N^b\right] =&
 \frac{i }{2 }  \tilde\cT^{ab} 
 +\frac{i }{2 } c g^{ab} 
-\frac{i }{2 } g^{ab} \cR^2
+i \left( X_N^a X_n^b - X_n^a X_N^b \right) \; , \\
\left[T^a,X_n^b\right] =&
  \frac{i }{2  }  \tilde\cT^{ab} 
 + \frac{i }{2 }c  g^{ab}
-\frac{i }{2}  g^{ab} \cR^2 
- i \left( X_N^a X_n^b - X_n^a X_N^b \right) \; .
\end{align}
\end{subequations}
We also note the identities
\begin{equation}
\begin{aligned}
 X_N^a X_n^b - X_n^a X_N^b &= X_n^b X_N^a -X_N^b X_n^a = \det{A} \left(\tilde 
X^a X^b - X^a \tilde X^b \right) \; , \\
 [X^a,\tilde X^b] &= i \tilde\cM^{ab} \; ,
\end{aligned}
\end{equation}
which follow from \eqref{XNXn-CR}.
One can check that the commutation relations \eqref{eq:commutators_XNXn} reduce 
to the Poisson brackets 
\eqref{Poisson-xxt} in the semi-classical 
(large $N$, $n$) limit.
\paragraph{\texorpdfstring{$X,Y,T$}{X,Y,T} variables.}
We note that the following  combinations play a special role:
\begin{equation}
\label{eq:def_lin_combi}
\begin{aligned}
X^a &= X^a_N + X^a_n \; , \\
Y^a &\coloneqq X_N^a  - X_n^a
 = \frac {-2}{\det \tilde A} \big(2(c+2)X^a +  \tilde X^a\big) \;, \\
 Y^{ab} &\coloneqq \cM_N^{ab}  - \cM_n^{ab} \; .
\end{aligned}
\end{equation}
$Y^a$ is the fuzzy counterpart of $y^a$, as defined in \eqref{y-def}, and it 
satisfies the same simple relations
\begin{equation}
\begin{aligned}
 \left[T^a,Y^b\right]
   &= i(Y^a X^b - X^a Y^b) \; ,\\
 \left[\cR^2,Y^b\right] &= 0 \; .
\end{aligned}
\end{equation}
This means that $Y^a$ could serve as a definition of a fuzzy 4-sphere based on 
$SO(5)$
rather than $SO(6)$, which respects $\cR^2 = \mathrm{const}$.
However, this would again remove the ``momentum'' degrees of freedom $T^a$ 
required for the mechanism 
for gravity of \cite{Steinacker:2016vgf},
and we will not pursue this possibility any further here.
However, we  note  
\begin{align}
 [Y^a, Y^b] &= \frac{2i}{\det{\tilde{A}} }
 Y^{ab} (-2c + \cR^2)
- \frac{2i }{\det{\tilde{A}}}
\left(Y^a T^b - T^a Y^b \right) 
 - \frac{2}{ \det{\tilde{A}}  } 
\left(Y^a X^b - X^a Y^b \right) 
\end{align}
and 
\begin{align}
 \left[T^a, T^b\right] &=
 \ \frac{i }{2 }\det{\tilde{A}}\ Y^{ab}+ i \cM^{ab}(2c - \cR^2)
 + \tilde\cT^{ab} - g^{ab} \cR^2  
 -i \left( X^a T^b - T^a X^b \right) \nn\\
&= -i\tilde \cM^{ab}-i\cM^{ab}(\cR^2+4)
+ \tilde\cT^{ab} - g^{ab} \cR^2  
 -i \left( X^a T^b - T^a X^b \right) \; .
\end{align}
This is consistent with the commutative limit \eqref{Poisson-xxt}, because 
\begin{align}
  \tilde\cT^{ab} - g^{ab} \cR^2  
 -i \left( X^a T^b - T^a X^b \right) = -\frac i2(X^a T^b - T^a X^b  + h.c.) \ .
\end{align}
Moreover, we verify agreement with \eqref{scalar-products-class} explicitly via 
\begin{equation}
\begin{aligned}
T\cdot T&=
\frac{1}{16} \left(4 N^2 n^2  + 12(Nn-1) (N+n)
+N^2 +n^2 +26N n 
\right) \one
-4 \Delta^2
+6 \Delta   \\ 
&\approx \frac{N^2 n^2}{4} - 4 \delta^2  \; ,
\end{aligned}
\end{equation}
which follows as $\Delta \approx \delta$ due to the definitions 
\eqref{scalar-products-class} and \eqref{inner-products-Nn}.
\subsection{Summary of commutation relations}
For convenience, we collect the most transparent form of the commutation 
relations for the vector generators of fuzzy $\cS^4_{\L}$:
\begin{equation}
\label{commutators-summary}
\begin{aligned}
 [X^a, X^b] &= i \cM^{ab}  \; ,\\ 
 [X^a, Y^b] &=  i Y^{ab} = i (\cM_N^{ab} - \cM_n^{ab})  \; ,\\ 
 [X^a, T^b] &=  - i \tilde \cT^{ab} + i(\cR^2 - c) g^{ab}  \; , \\ 
 [Y^a, Y^b] &=  \frac{2i}{\det{\tilde{A}} } 
  \Big( Y^{ab} (-2c + \cR^2) -Y^a T^b + T^a Y^b 
 +i \left(Y^a X^b - X^a Y^b \right) \Big)  \; ,\\
\left[T^a, T^b\right] &=
  \frac{i }{2 }\det{\tilde{A}}\, Y^{ab}+ i (2c - \cR^2)\cM^{ab}
-i \left( X^a T^b - T^a X^b \right)
 + \tilde\cT^{ab} - g^{ab} \cR^2  \; , \\
\left[T^a,Y^b\right] 
   &=  i(Y^a X^b - X^a Y^b) \; ,
\end{aligned}
\end{equation}
where $a,b = 1,\ldots,5$. The Poisson brackets \eqref{Poisson-brackets-collect} 
are recovered in the semi-classical limit  by 
replacing $[\cdot,\cdot] \to i \{\cdot,\cdot\}$ and dropping sub-leading terms.
%
%
%
\section{Embeddings in matrix models}
\label{sec:M-M-embeddings}
The main motivation of all these consideration is to find solutions of these 
generalized fuzzy spheres
-- possibly with extra dimensions -- in Yang-Mills matrix models, and in 
particular the IKKT model.
These models are defined by the action 
\begin{align}
  S_{\rm YM}[\cY] &= \frac 1{g^2}\Tr \Big(-[\cY_A,\cY_B][\cY^A,\cY^B]\, + \mu^2 
\cY^A \cY_A \Big) \ .
 \label{bosonic-action}
\end{align}
Here $\cY^A, \ A=1,\ldots,10$, are hermitian matrices, for which indices are 
raised 
and lowered with $\d^{AB}$. 
The parameter $\mu^2$ introduces a scale\footnote{In the Minkowski case, such 
mass terms are
effectively introduced as IR regulators, both for space-like and time-like 
matrices \cite{Kim:2011cr,Kim:2012mw}.}. 
The equations of motion are 
\begin{align}
 \Box_\cY \cY^A + \mu^2 \cY^A = 0 \; , \text{ with} \quad \Box_{\cY} \cY^A 
\coloneqq \left[\cY_B \left[ \cY^B, \cY^A \right] \right] \; .
\label{eq:def_Laplacian}
\end{align}
This (Euclidean) model does not have any non-trivial solutions for $\mu^2 > 0$, 
but it does have many solutions for $\mu^2<0$. 
Although the latter case is  unstable,  this may be justified by starting with a 
bare ``mass'' $\mu^2 > 0$ and taking into account quantum corrections. 
Computing, for example, 1-loop corrections around some given background 
$\cY^A$, such as fuzzy $S^4$, one obtains an
effective action 
\begin{align}
  S_{\rm eff}[\cY] =   S_{\rm YM}[\cY] + S_{\rm 1-loop}[\cY]\,.
\end{align}
The full form of $S_{\rm 1-loop}(\cY)$ is clearly very complicated. 
On  backgrounds which respect some global symmetry, 
 $S_{\rm 1-loop}(\cY)$ will respect that symmetry.
Consider, for instance, backgrounds of the form (c.f. 
\eqref{eq:ansatz_phasespace_embed})
\begin{align}
  {\cY}^A   =  \begin{pmatrix}
                    \cY^a_{(1)}  \\ \cY^a_{(2)} 
                     \end{pmatrix}
  =  \begin{pmatrix}
                    r_1  X^a_{N} +  r_2 X^a_{n}  \\ r_3 T^a
                     \end{pmatrix} \, , 
                     \quad a=1,\ldots,5 \; ,
\end{align}
which preserve $SO(5)$. Define the $SO(5)$-invariant observables
\begin{align}
 \cR_{i}^2 &= \cY_{(i)} \cdot \cY_{(i)} \ , \qquad 2\cR_{12} = \cY_{(1)} \cdot 
\cY_{(2)} + h.c. \ .
\end{align}
Then the dependence of the background on $r_{i}$ 
is captured in $V_{\rm 1-loop}(r_i) = S_{\rm 1-loop}[{\cY}(r_i)]$, which 
yields the effective matrix model 
\begin{subequations}
\begin{align}
 S_{\rm eff}[\cY] &=  \frac 1{g^2}\Tr \Big(-[\cY_A,\cY_B][\cY^A,\cY^B]\, +  
V_{\rm eff}[\cY]\Big) \, , \\
  V_{\rm eff}[\cY] &= \mu^2(\cR_{(1)}^2 + \cR_{(2)}^2) +  V_{\rm 
1-loop}\big(\cR_{(i)}^2,\cR_{(12)}\big) \, .
 \end{align}
 \end{subequations}
Now assume that $S_{\rm eff}$ has a non-trivial minimum as a function of $r_i$;
this happens, for example, in the IKKT model for the basic $S^4_N$ 
\cite{Steinacker:2015dra}.
We can then expand $V_{\rm 1-loop}(\cR_{(i)}^2,\cR_{(12)})$ up to quartic order 
around these background values,
and rewrite it in terms of the above observables $\cR_i^2$ etc.
Then this background is a solution of an effective matrix model of the form
\begin{equation}
\label{S-eff-MM}
\begin{aligned}
  S_{\rm eff}[\cY] =&  \frac 1{g^2}\Tr \Big(-[\cY_A,\cY_B][\cY^A,\cY^B]\, 
  + \mu_1^2\cR_{(1)}^2 + \mu_2^2\cR_{(2)}^2  \\
  &\phantom{\frac 1{g^2}\Tr \Big(} + \l_1 (\cR_{(1)}^2)^2 + \l_2 
(\cR_{(2)}^2)^2 + \mu_{12} \cR_{(12)} + \ldots  \Big)
\end{aligned}
\end{equation}
dropping a constant and some higher-order terms for simplicity.
Now $\mu_i^2$ can be negative, while the quartic potential should be positive 
definite. 
This leads to the following equations of motion
\begin{equation}
\begin{aligned}
  \Box_\cY \cY^a_{(1)} &= - \frac {\mu_1^2}2\, \cY^a_{(1)} - \frac{\mu_{12}}2\, 
\cY^a_{(2)} - \{\l_1 \cR_1^2,\cY^a_{(1)}\}_+  \; ,\\
  \Box_\cY \cY^a_{(2)} &= - \frac{\mu_2^2}2\,  \cY^a_{(2)} - \frac{\mu_{12}}2\, 
\cY^a_{(1)} -\{\l_2 \cR_2^2,\cY^a_{(2)}\}_+ \; ,
\end{aligned}
\end{equation}
where $\{\cdot,\cdot\}_+$ denotes the anti-commutator.
These are the equations we will solve in this paper.
In fact, we will mostly drop also $\mu_{12}$ for simplicity; this means no loss 
of generality for the 
phase-space solutions in Section \ref{sec:phase-space-embed} where $\cR_{12} = 
0$, but it is a non-trivial restriction for the 
 solutions in Section \ref{sec:spherical-solutions}. A more complete treatment 
of the latter should be given elsewhere.

To find solutions of these equations, we will compute the matrix 
Laplacians $\Box_{\cY} \cY$
for all possible choices $\cY= X_N, X_n, T$, based on 
the commutator relations \eqref{eq:commutators_XNXn} and the 
multiplication algebra \eqref{MNnO-algebra}.
Some explicit formulae are delegated to
Appendix \ref{sec:laplacians}.
%
%
\subsection{Spherical embedding}
\label{sec:spherical-solutions}
As a first case study, we investigate an embedding of the form
\begin{align}
 {\cY}^A  = \begin{pmatrix}
        a X_N^a + b X_n^a \\
        g \, X_N^a + h \, X_n^a
       \end{pmatrix} 
=\begin{pmatrix}
        a &  b  \\
        g & h 
       \end{pmatrix}
      \begin{pmatrix}
        X_N^a \\
         X_n^a
       \end{pmatrix}
       \label{eqn:ansatz_spherical_embed}
\end{align}
for $a,b,g,h \in \R$. 
If the $2\times 2$ matrix 
$\scriptsize\begin{pmatrix}
                            a &  b  \\  g & h 
                           \end{pmatrix}$
is invertible, such a background can be interpreted as 
$S^4_N\times S^4_n$, possibly sheared. If the matrix is degenerate, then
one can  find a rotation such that one of the lines in 
\eqref{eqn:ansatz_spherical_embed} vanishes.

For this ansatz \eqref{eqn:ansatz_spherical_embed},
the full matrix Laplacian is explicitly decomposed as 
\begin{align}
 \Box \equiv (a^2 + g^2) \Box_{X_N} + (b^2 +h^2)\Box_{X_n} + (a 
b + g h) \Box_{mix} \; ,
\end{align}
where we define the mixed Laplacian as
\begin{align}
 \Box_{mix} \coloneqq \sum_{a=1}^5 \left(  \left[X_N^a , \left[X_n^a, \,\cdot \,
\right]\right] + \left[X_n^a , \left[X_N^a, \,\cdot \, \right]\right] \right) 
\; .
\label{mixed-laplace}
\end{align}
The explicit results for the various contributions are given in Appendix 
\ref{sec:laplacians}. For the 
generic case, the action of the Laplacian leads to
\begin{subequations}
\begin{align}
 a \Box X_N + b \Box X_n 
=& F_{X_N} X_N + F_{X_n} X_n + F_{X_N,\cR^2} \left\{ X_N,\cR^2 \right\}_+
+ F_{X_n,\cR^2} \left\{ X_n,\cR^2 \right\}_+ \; ,\\
 g \Box X_N + h \Box X_n 
=& f_{X_N} X_N + f_{X_n} X_n + f_{X_N,\cR^2} \left\{ X_N,\cR^2 \right\}_+
+ f_{X_n,\cR^2} \left\{ X_n,\cR^2 \right\}_+ \; .
\end{align}
\end{subequations}
Next, we impose the following conditions to obtain a solution for the 
matrix model  with $\mu_{12} = 0$ in \eqref{S-eff-MM} (i.e.\ without mixing 
term $\cY_{(1)} \cdot \cY_{(2)}$ in the effective action):
\begin{subequations}
\begin{alignat}{3}
 \frac{ F_{X_N}}{a} &= \frac{ F_{X_n}}{b} 
 \qquad & &\text{and} \qquad &
\frac{ f_{X_N}}{g} &= \frac{ f_{X_n}}{h} \; ,
\label{eq:condition_XNXn_TwoX} \\
 \frac{ F_{X_N,\cR^2}}{a} &= \frac{ F_{X_n,\cR^2}}{b} 
 \qquad & &\text{and} \qquad &
\frac{ f_{X_N,\cR^2}}{g} &= \frac{ f_{X_n,\cR^2}}{h}  \; .
\label{eq:condition_noRR_TwoX}
\end{alignat}
\end{subequations}
We start with condition \eqref{eq:condition_noRR_TwoX}, which yields six 
solutions
\begin{subequations}
\begin{align}
 \{b =&  a  \quad \text{and} \quad  h = g \}
 \; , \qquad 
 \{ b = - a  \quad \text{and} \quad  h = - g\}  \; , 
 \label{eqn:sol_noRR_twoX_1}\\
\{ b =&  a  \quad \text{and} \quad  h = - g \}  
\; , \qquad 
\{b = - a  \quad \text{and} \quad  h =  g \} \; ,
\label{eqn:sol_noRR_twoX_2}\\
\bigg\{ g =& \frac{R^{+}}{(N-n)^2 - 6  \det{\tilde{A}}} \bigg\}
\; , \qquad 
\bigg\{ g = \frac{R^{-}}{(N-n)^2 - 6  \det{\tilde{A}}} \bigg\}
\; ,\quad \text{with}  
\label{eqn:sol_noRR_twoX_3}\\
R^{\pm}=& h ( N-n)^2 
\ \textcolor{red}{\pm} \Bigg(  h^2  (N-n)^4
- \left(  (N-n)^2-6 \det{\tilde{A}}\right) 
\bigg[
\left( 6 \det{\tilde{A}} + (N-n)^2 \right)h
\notag \\
&+(a-b)^2 (N-n)^2 + 6 \det{\tilde{A}} (b^2 -a^2)
\bigg] 
\Bigg)^{\frac{1}{2}}
\nn \; .
\notag
\end{align}
\end{subequations}
We proceed by imposing on each solution of \eqref{eq:condition_noRR_TwoX} the 
remaining constraint \eqref{eq:condition_XNXn_TwoX}.
\paragraph{Two archetypal solutions}
As it turns out, the four solutions  \eqref{eqn:sol_noRR_twoX_1}, 
\eqref{eqn:sol_noRR_twoX_2} are variations of two archetypal solutions which we 
construct from a simplified ansatz 
\begin{align}
 {\cY}^A = \begin{pmatrix}
        a X_N^a + b X_n^a \\   0
       \end{pmatrix}
       \label{eq:ansatz_simple_embed} \, ,
\end{align}
 with $g=h=0$ right from the beginning. For the constraint
\begin{align}
 \frac{ F_{X_N,\cR^2}}{a} &= \frac{ F_{X_n,\cR^2}}{b}  \; ,
\end{align}
we find three solutions
\begin{align}
 b = \pm a \; , \qquad b= \frac{(N-n)^2  - 6 \det{\tilde{A}} 
}{(N-n)^2  + 6 \det{\tilde{A}}} a \; .
\end{align}
However, only $b= \pm a$ is compatible with 
\begin{align}
  \frac{ F_{X_N}}{a} &= \frac{ F_{X_n}}{b} \; ;
\end{align}
while for the third solution this constraint imposes $a=0$. Therefore, the 
simple ansatz \eqref{eq:ansatz_simple_embed} has precisely two non-trivial 
solutions 
\begin{align}
 \Box (X_N +X_n) &= \lambda \;  (X_N +X_n) \; , \qquad
\lambda = 4  \; ,
\label{eqn:sol_spherical_XN+Xn}
\end{align}
and
\begin{equation}
\label{eqn:sol_spherical_XN-Xn}
\begin{aligned}
 \Box (X_N -X_n) &= \lambda_1  \; (X_N -X_n) + \lambda_2 
\; \left\{ \cR^2, (X_N -X_n) \right\}_+  \\[1ex]
 \lambda_1 &=2 \ \frac{3 N^2 +12 N + 3 n^2 +12 n +2N n+8
}{(N+n+4)(N+n+2)}   \\
 \lambda_2 &=- \frac{4}{(N+n+4)(N+n+2)}  \; . 
\end{aligned}
\end{equation}
Remarkably, we recover precisely the $X = X_N+X_n$ and $Y = X_N-X_n$ generators.
Clearly, rescalings of the form $(X_N \pm X_n) \to a\, (X_N \pm X_n)$ and $\l_i 
\to a^2 \l_i$ provide the solutions for arbitrary $a\in \R$.
Moreover, the eigenvalue equation \eqref{eqn:sol_spherical_XN+Xn} agrees with 
the semi-classical computation \eqref{eq:Laplace_eigenvalue_x_classic}.

Now we proceed to the generic cases.
\paragraph{Case: $b=a$ and $h=g$}
The additional constraint \eqref{eq:condition_XNXn_TwoX} is automatically 
satisfied for any $a$ and any $g$. Thus, this solution behaves as two copies 
$a(X_N+X_n)$, $g(X_N+X_n)$ of \eqref{eqn:sol_spherical_XN+Xn} with a common 
eigenvalue $\lambda=4(a^2+g^2)$.
\paragraph{Case: $b=-a$ and $h=-g$}
We find that \eqref{eq:condition_XNXn_TwoX} is automatically satisfied for 
any $a$ and any $g$. The solution behaves as two copies 
$a(X_N-X_n)$, $g(X_N-X_n)$ of \eqref{eqn:sol_spherical_XN-Xn} with a common 
eigenvalues $\lambda_1$ , $\lambda_2$, in which one replaces $a^2$ by $a^2+g^2$.
\paragraph{Case: $b=a$ and $h=-g$}
It turns out that \eqref{eq:condition_XNXn_TwoX} only holds if $g=0$, while 
$a$ remains unconstrained. The solution is identical to 
\eqref{eqn:sol_spherical_XN+Xn}.
\paragraph{Case: $b=-a$ and $h=g$}
Again, imposing \eqref{eq:condition_XNXn_TwoX} forces $a=0$, but $g$ 
remains arbitrary. The solution is identical to \eqref{eqn:sol_spherical_XN+Xn} 
for $a$ replaced by $g$.
\paragraph{Remark}
There are two additional solutions \eqref{eqn:sol_noRR_twoX_3} for the first 
condition \eqref{eq:condition_noRR_TwoX}, which might be of potential interest 
because the anti-commutator contributions involving $\cR^2$ are vanishing in 
these cases. This is a surprising feature of \eqref{eqn:sol_noRR_twoX_3} which 
one can explicitly verify by inserting the expression into the $\cR^2$ 
anti-commutators. However, we are currently unable to verify the compatibility 
with the remaining condition \eqref{eq:condition_XNXn_TwoX} analytically. We 
have, however, verified the existence of numerical solutions to 
\eqref{eq:condition_XNXn_TwoX} for a number of explicit choices of $N\gg n>0$.
\paragraph{Discussion}
It is interesting to see that the explicit solutions found above are  of the 
form 
\begin{align}
 X_N + X_n &= X \; , \qquad 
X_N - X_n = Y 
\; .
\end{align}
Recall that $Y$ was identified in \eqref{eq:def_lin_combi} 
as generators which commute with $\cR^2$. We also recovered the fact that  $X= 
X_N+X_n$ 
is an eigenvector of the corresponding matrix 
Laplacian without any $\cR^2$-contributions. 
We have found numerical evidence for additional solutions 
other than $X_N \pm X_n$, which arise from \eqref{eq:condition_noRR_TwoX}. 
\subsection{Phase-space embedding}
\label{sec:phase-space-embed}
Second, we consider an embedding ansatz of the form
\begin{align}
 {\cY}^A = \begin{pmatrix}
        a X_N^a + b X_n^a \\
        d\, T^a 
       \end{pmatrix} \; .
       \label{eq:ansatz_phasespace_embed}
\end{align}
This is perhaps the most interesting case, because the $T^a$ plays the role of momentum 
generators for $S^4$, 
which has important consequences for the metric fluctuations on such a 
background.
Moreover, the embedding of $T^a$ amounts to a squashed embedding of $\C P^2$ (see Section \ref{sec:global}), 
which is
expected to entail a low-energy physics with 3 generations. 

There are two interesting special cases:
\begin{itemize}
 \item 
 Only $X^a$ is embedded  ($b=a$, $d=0$). This is the case studied in 
\cite{Steinacker:2016vgf}.
 
 \item 
  Only $T^a$ are embedded ($a=b=0$). This is related to the momentum 
interpretation of the matrix model \cite{Hanada:2005vr}, which was argued to 
lead to the Einstein equations.
\end{itemize}

The Laplacian acts on the first components of 
\eqref{eq:ansatz_phasespace_embed} via 
\begin{align}
\Box (a X_N^a + b X_n^a) \equiv  F_{X_N} X_N + F_{X_n} X_n + F_{X_N,\cR^2} 
\left\{ X_N,\cR^2 \right\}_+
+ F_{X_n,\cR^2} \left\{ X_n,\cR^2 \right\}_+ \; ;
\end{align}
whereas the Laplacian on the $T^a$ component is much simpler, c.f.\ Appendix 
\ref{sec:laplacians}
\begin{equation}
\label{eq:phase_space_T}
\begin{aligned}
 \Box (d \ T^a)=&  \ \frac{1}{2}  \bigg(
   (5 a^2   +6  a  b +5 b^2 ) 
 +(9 N^2 +30 N + 9 n^2 +30  n +2 N n ) d^2
\bigg) d\ T^a \\
&- 6 d^2 \left\{\cR^2, d\ T^a\right\}_+\; .
\end{aligned}
\end{equation}
As in the previous embedding scenario, we impose the following two conditions 
to determine the variables in the ansatz:
\begin{subequations}
\begin{align}
 \frac{ F_{X_N,\cR^2}}{a} &= \frac{ F_{X_n,\cR^2}}{b}  \; ,
\label{eq:RR-constraint} \\
 \frac{ F_{X_N}}{a} &= \frac{ F_{X_n}}{b} \; .
 \label{eq:XNXn-constraint}
\end{align}
\end{subequations}
One readily obtains four solutions to the constraint \eqref{eq:RR-constraint}
\begin{subequations}
\begin{align}
 \big\{ b &= \pm a \big\} \; , \\
\bigg\{ d &= \pm \frac{ \sqrt{a-b} }{2 \sqrt{3} }
\sqrt{ 
\frac{6 \det{\tilde{A}} (a+ b) -  (N-n)^2 
(a-b)}{ \det{\tilde{A}}^2 } 
} \bigg\} \label{eq:sol_CC}
\end{align}
\end{subequations}
Next, we analyze the compatibility of each solution with the remaining 
constraint \eqref{eq:XNXn-constraint}.
\paragraph{Case: $b=a\neq0$}
The remaining condition \eqref{eq:XNXn-constraint} is solved for $
 d =0$, but arbitrary $a \in \R$.
In other words, the solution reduces exactly to \eqref{eqn:sol_spherical_XN+Xn}.
\paragraph{Case: $b=-a\neq0$}
In this case, the condition \eqref{eq:XNXn-constraint} does not restrict the 
parameters at all, such that $a, d \in \mathbb{R} $ are arbitrary. Then the 
solution is
\begin{equation}
\label{T-Y-embedding}
\begin{aligned}
 \Box \begin{pmatrix} a (X_N -  X_n) \\ d\, T  \end{pmatrix}
&= \begin{pmatrix} \lambda_1 \, a ( X_N  - X_n ) \\ \lambda_2 \, d\, T
\end{pmatrix}  
+
\begin{pmatrix} \lambda_3 \,  \left\{\cR^2,  a ( X_N  - X_n ) \right\}_+ \\  
\lambda_4 \,  \left\{\cR^2,  d\, T \right\}_+
\end{pmatrix}\; ,   
 \\[1ex]
\lambda_1 &= 2   \frac{3 N^2 +3 n^2 +12 N +12 n +2Nn +8 }{(N+n+4)(N+n+2)} a^2
 \\
&\qquad+\frac{1}{2} d^2 \left(3 N^2+3 n^2 +12 N +12 n + 2 Nn\right) \; , 
\\
\lambda_2 &= 2 a^2 
+\frac{1}{2} d^2 \left(9 N^2 +9 n^2 +30 N +30 n +2 N n \right) \; , 
\\
\lambda_3 &= - \left( \frac{4 a^2}{(N+n+4)(N+n+2)}+3 
d^2\right) \; , \\
\lambda_4 &=-6 d^2 \; .
\end{aligned}
\end{equation}
%
\paragraph{Case: $a=b=0$}
Here, there is nothing to prove. The valid solution is
\begin{equation}
\begin{aligned}
 \Box \begin{pmatrix} 0 \\ \, T  \end{pmatrix}
&= \lambda_1 \, \begin{pmatrix} 0 \\   T \end{pmatrix} +
\lambda_2 \, \begin{pmatrix} 0 \\ \left\{\cR^2, \, T \right\}_+ \end{pmatrix} 
\; , \\ 
\lambda_1 &=\frac{1}{2}(9 N^2 +30 N + 9 n^2 +30  n +2 N n )  \; , \\
\lambda_2 &=-6  \; .
\end{aligned}
\end{equation}
Clearly, rescaling $T \to d\, T$ and $\l_i \to d^2 \l_i$ provides the solution 
for arbitrary $d\in \R$.
This solution for $T^a$ is consistent with the semi-classical computation
\eqref{eq:Laplace_eigenvalue_t_classic} for $t^a$.
\paragraph{Case: non-trivial $d = \pm \ldots$}
Starting from the solution \eqref{eq:sol_CC} of \eqref{eq:RR-constraint},
 we find the following four solutions for \eqref{eq:XNXn-constraint}:
\begin{subequations}
\begin{align}
 \{ b =& \pm a  \} \qquad \text{and} \\
 \bigg\{ b =& \frac{R^{\pm}}{P} \bigg\} \qquad \text{with} \label{eq:sol_BB} 
\\
R^{\pm}=&-a (N-n)^2 \left(21 N^2 +21 n^2 +150 N +150 n +58 n N + 216\right) 
\notag 
\\
&\textcolor{red}{\pm} 2 \sqrt{3} \left|a \det{\tilde{A}}\right| 
\bigg(
147 N^4 +147 n^4
+924 N^3 +924 n^3
+1452 N^2 +1452 n^2 \notag \\
&\qquad 
+28 n N^3 +28 n^3 N
-780 n N^2 -780 n^2 N
-302 n^2 N^2
-2472 n N
\bigg)^{\frac{1}{2}}
\notag \\
P=&
6  \det{\tilde{A}}  \left(7 N^2 + 7  n^2 +6 N +6 n -60  n N\right) \notag 
\\
& +21 N^4  +21 n^4 
+102 N^3 +102 n^3 
+120 N^2 +120 n^2 \notag\\
&-16 n N^3 -16 n^3 N
-102 n N^2 -102 n^2 N
-10 n^2 N^2 -240 n N
\notag
\end{align}
\end{subequations}
Let us discuss the various possible solutions:
\begin{itemize}
 \item Choosing \eqref{eq:sol_CC} together with $b=a$ implies that $d=0$, 
i.e. the $T$-direction disappears, and the $\cR^2$-terms vanish. The solution 
is then again given by \eqref{eqn:sol_spherical_XN+Xn}.
%
\item Choosing \eqref{eq:sol_CC} together with $b=-a$ yields
\begin{align}
 d = \pm \sqrt{\frac{  -  (N-n)^2 a^2 }{ 3 }} \in i \mathbb{R} \; ,
\end{align}
which is not a viable solution.
\item Choosing \eqref{eq:sol_CC} together with \eqref{eq:sol_BB} yields two
non-trivial $d$ together with vanishing $\cR^2$-terms, i.e.
\begin{align}
 &\Box \begin{pmatrix} a \, X_N + b(a)\, X_n \\ d(a) T  
\end{pmatrix}
= \begin{pmatrix} \lambda_1^{\pm}\,  (a\, X_N + b(a)\, X_n) \\ 
\lambda_2^{\pm}\, 
d(a)\, T  \end{pmatrix} 
- 6 d^2(a)\begin{pmatrix} 0 \\ 
\left\{\cR^2, d(a)\, T \right\}_+ \end{pmatrix} 
\label{eq:phase-space_tricky_case}
\end{align}
The precise expressions for $\lambda_1^{\pm}, 
\lambda_2^{\pm}$ and $b(a), d(a)$ can be readily obtained by inserting 
\eqref{eq:sol_CC} and
\eqref{eq:sol_BB}. Due to the elaborate and somehow uninstructive nature of 
these expressions we refrain from presenting them here.

\end{itemize}
\paragraph{Discussion}
Again, we found solutions given by the simple linear combinations $X_N \pm X_n$. 
In particular, we found an explicit, simple solution of
type $(Y,T)$  \eqref{T-Y-embedding}, which could be naturally interpreted as 
fuzzy 4-sphere
with self-intersecting fuzzy extra dimensions. This is exactly what we were looking for.
This solution (or ansatz) represents arguably the most interesting and sophisticated
candidate to obtain 4-dimensional physics from the (Euclidean) IKKT model, and it
 involves all 10 matrices of this model.
However, we cannot make any statements about the scale parameters at this point; 
 determining these
would  require at least a 1-loop computation.  

Another, rather special solution, is obtained by solely embedding $T$. 
This should provide a well-defined, finite-dimensional realization of the 
momentum picture put forward in \cite{Hanada:2005vr},
which should allow to clarify the mechanism for gravity in this scenario.
Lastly, we found a solution with non-trivial $T$ embedding and partially vanishing $\cR^2$ 
contributions; 
however, the classical analog, if any, is not obvious.  
%
%
%
\section{Effective metric and prospects for 4D physics}
\label{sec:eff_metric_4d_physics}
The effective metric for fluctuation modes on these backgrounds is extracted 
from the kinetic term in the matrix model, i.e. from the matrix Laplacian $\Box 
= [\cY^A,[\cY_A,\cdot]]$.
In the semi-classical limit, this defines a second order differential operator 
which encodes 
the effective metric on $\cO_\L$. As shown in \cite{Steinacker:2010rh}, this is 
indeed the Laplacian for some effective 
metric on $\cO_\L$, which is in general distinct from the induced (pull-back) 
metric.
The most transparent way to extract this metric in the semi-classical case is 
as follows: From the kinetic term for some fluctuation field $\phi \in 
\End(\cH)$ 
\begin{align}
 -[\cY^A,\phi][\cY_A,\phi] \ &= \ D^A \phi  D_A \phi \sim  G_\cY^{\mu\nu} 
\del_\mu \phi  \del_{\nu} \phi \;  , \quad \text{ with}
   \qquad  D^A \coloneqq -i[\cY^A,\cdot] 
 \label{kinetic-term}
\end{align}
we can read off the effective  metric in the matrix model 
(up to a possible conformal factor, cf. \cite{Steinacker:2010rh,Steinacker:2016vgf}) 
\begin{align} 
 \  G_\cY^{\mu\nu} =  \{\cY^A,\xi^\mu\}\{\cY_A,\xi^\nu\} \  =  \ D^A \xi^\mu  
D_A \xi^\nu \ .
 \label{metric-vielbein-gen}
\end{align}
Here $\xi^\mu$ denotes some local coordinates on $\cO_\L$.
The effective metric \eqref{metric-vielbein-gen} can be expressed in terms of 
an (generalized) frame
\begin{align} 
G_\cY^{\mu\nu} = e^{A\mu}e_A^{\ \nu}, \qquad 
 \  e^A = \{\cY^A,\cdot \} = e^{A\mu}\del_\mu  \ \  = D^A  \ .
 \label{vielbein-e}
\end{align}
For the coordinates $\xi^\mu= (x^a,y^a,t^a)$ and the embedding $\cY^A$, we denote
\begin{subequations}
\label{eq:notation_metric}
\begin{align}
 \left( G_{\cY}^{\mu\nu} \right) &= \begin{pmatrix}
 G(x,x) & G(x,y) & G(x,t) \\
 G(y,x) & G(y,y) & G(y,t)\\
 G(t,x) & G(t,y) & G(t,t)            
           \end{pmatrix} \\
           \mathrm{with } \quad
 G(x^b,x^c) &\equiv G^{bc}(x,x) = \{\cY^A,x^b\} \{\cY_A,x^c\}           
 \quad \mathrm{etc.}
\end{align}
\end{subequations}
If only $X^a$ is embedded, this gives 
\begin{align}
 G_X^{\mu\nu} = g_{ab} \{x^a,\xi^\mu\} \{x^b,\xi^\nu\} \,. 
 \label{metric_embed_x}
\end{align}
Note that this is a metric on the entire orbit $\cO_\L$. 
For scalar functions on the basic $\cS^4_N$, which depend only on 
$x^a$, \eqref{metric_embed_x} reduces to the round metric on $S^4$, as 
discussed 
in \cite{Steinacker:2016vgf}.

To compute such metrics  on $\cO_\L$, one can use the local description of 
$\tilde\cO_\L$ provided by \eqref{O-SSS-map}.
We will illustrate this first by computing
the pull-back metric $g$ on  $\tilde\cO_\L \hookrightarrow S^4\times S^4 \times 
\R^5$.
Since this is a homogeneous space, it suffices to evaluate the metric at any 
given reference point  
$p \in \tilde\cO_\L$, and use $x_N^{1,\ldots,4}, x_n^{1,2,3}, t^{1,2}$  and 
$\vartheta$ as local coordinates as in \eqref{ref-point-Otilde}.
Then the induced metric at $p$ is 
\begin{equation}
d s^2_g\big|_{p} =  R_N^2 \left( \sum_{\mu=1}^4 d x_N^\mu x_N^\mu \right) 
  + R_n^2 \sin^2\vartheta \left(\sum_{\mu=1}^3 d x_n^\mu x_n^\mu \right)  
   + R_t^2 \left(\sum_{\mu=1}^2 d t^\mu d t^\mu\right) 
    + R_N^2 d\vartheta  d\vartheta \ .
\nonumber
\end{equation}
The effective metric will generically contain additional contributions from the 
tensor $t^{ab}$.
Depending on the embedding, it is possible that one sphere is effectively very 
large and describes ``space-time'', while the remaining spheres  
have either a large mass gap or  some are degenerate. Then
 an effectively 4-dimensional theory would arise at low energies.
\subsection{Momentum space embeddings}

Consider first the momentum space embeddings of the form $ \cY^A = ( T^a , 0 
)^T$.
The effective metric \eqref{vielbein-e} on $S^4$ then takes a somewhat more 
complicated form, which is obtained  using  the Poisson brackets 
\eqref{Poisson-brackets-collect}
(in the notation of \eqref{eq:notation_metric})
\begin{align}
 G^{bc}(x,x) 
 = \left(\frac{N^2+n^2}{2} -2\cR^2 
\right) 
t^{bc} - t^b t^c
+\left(\cR^4 - \frac{(N^2-n^2)^2}{16} \right)g^{bc}\,,
\label{eff-metric-T-embed-full}
\end{align}
for $b,c=1,\ldots,5$. 
This is not equal to $g^{ab}$, and it is not constant on $\cK$ since $t^{ab}$ 
decomposes as $(2,0) \oplus (0,0)$ under $SO(5)$. 
However, for  $N\gg n\geq1$ we can approximate
\begin{align}
 t^{ab} \approx \frac{N^2}{4} g^{bc} \; ,\qquad 
x^a \approx x_N^a\approx y^a \; , \qquad m^{ab} \approx m_N^{ab} \approx  
y^{ab} \; , 
\label{eq:limit_metric}
\end{align}
and using $\cR^2 =\frac{1}{4} (N^2+n^2) + 2\delta $ for $\delta \in 
\left[ -\frac{Nn}{4},\frac{Nn}{4}\right]$ one observes
\begin{subequations}
\label{eq:TBack_all}
\begin{align}
 G^{bc}(x,x) = \left( \frac{N^2 n^2}{4} + 4 \delta^2+ O(N) 
\right) 
g^{bc}  -t^b t^c \sim O(N^2)\; .
\end{align}
The remaining metric components of $G_T^{\mu\nu}$ 
in this limit are as follows:
\begin{align} 
G^{bc}(t,t) &= \left(\frac{n^2 N^4}{16} +  N^2 \delta^2 + 
O(N^3)\right) g^{bc} \label{G_T(t,t)} \\
&\qquad + \left(N^2 n^2 +64\delta^2 +O(N) \right) x^b x^c 
- \left(4N^2 +O(N)\right) t^b t^c \sim O(N^4)\; ,
\nn \\
 G^{bc}(y,y) 
 &=n^2 x^b x^c \sim O(N^2) \; ,\\
 G^{bc}(x,t)  &=\left( \frac{N^2n^2}{4}+4\delta^2 
+O(N)\right) m^{bc}  \\
&\qquad 
+ \left(2\delta+O(n^2)\right) x^b t^c
-\left(6 \delta + O(n^2)\right) t^b x^c \sim O(N^3)\; , \nn
\\
 G^{bc}(x,y) &= -n^2 x^b x^c \sim O(N^2) \; ,\\
 G^{bc}(t,y) &=4 \delta \ t^b x_N^c \sim O(N^3) \;,
\end{align}
\end{subequations}
where $b,c=1,\ldots,5$.
The metric components along $y$-directions are highly degenerate, which reflects 
the fact that $\{t^\mu,y^\nu\} = 0$ \eqref{tmu-ynu}.
Hence, there is no propagation in $y$ space, which justifies dimensional 
reduction in  $y$ space. 
Moreover, we observe that the $G^{bc}(t,t) \sim O(N^4)$ dominates, which 
suggests that the harmonics in $t$ space acquire a large gap, while 
those in $y$ space decouple, leaving only $x$ space as physical space.

We also observe that for the above embeddings,  fluctuations of type 
\begin{align}
 \cY^a = \bar\cY^a + \d  \cY^a 
 = \big(\d^a_{\ b} + h^{a}_{\ b}(x)\big)\, T^b + \ldots
 \label{fluct-momentum-embedding}
\end{align}
amount to fluctuation of the vielbein. 
Hence, the corresponding metric fluctuations correspond to Goldstone 
bosons\footnote{See e.g. \cite{Kraus:2002sa} for a 
recent related discussion.}  of $SO(5)$.
A more detailed elaboration of these topics is postponed to future work.
\paragraph{4-dimensional momentum embedding.}
Finally, we observe that the following  background
\begin{align}
 \cY^\mu = P^\mu \coloneqq \cM^{\mu 5}, \qquad \mu = 1,\ldots,4
 \label{P-embedding}
\end{align}
also provides a solution, with
\begin{align}
 \Box_\cY \cY^\nu = 3 \cY^\nu \ .
\end{align}
Since $T^\mu = \cM^{\mu a} x_a \approx \cM^{\mu 5} \cR =  P^\mu  \cR$ at the 
north pole, this is closely related to the above 
embedding $\cY^a = T^a$, but it is algebraically simpler. 
Wave functions would still be\footnote{For the basic fuzzy sphere, there are no 
independent $P^\mu$ modes since 
$\cM^{ab}$ is tangential. Therefore the generalized $\cS^4_\L$ is essential.} 
$\phi(x)$ (for $N \gg n$), 
and perturbations of the vielbein would arise from (cf. \cite{Hanada:2005vr})
\begin{align}
 \cY^\mu = P^\mu + \d  \cY^\mu 
 = \big(\d^\mu_{\ \nu} + h^{\mu}_{\ \nu}(x)\big)\, P^\nu + \ldots \; .
 \label{fluct-P-embedding}
\end{align}
Although the $SO(5)$ symmetry is (mildly) broken to $SO(4)$ for this background, 
this may well be of physical interest, perhaps even more than the 
fully symmetric solution. The effective metric in this case is simply 
$G^{\mu\nu} = \d^{\mu\nu}$, 
corrected by  a conformal factor
which is not discussed here. This would thus lead to conformally flat metrics, 
which is very interesting from the 
cosmological point of view since FRW metrics are conformally flat. 
\paragraph{Gauge symmetry and diffeomorphisms.}
We briefly discuss gauge transformations 
for the above momentum embedding following \cite{Steinacker:2016vgf,Hanada:2005vr}.
At the north pole $ p \in S^4$ where $x^\mu = 0$, consider gauge transformations 
generated by
$\L = \xi_\mu(x) P^{\mu}$. They act on the background \eqref{P-embedding}
as follows
\begin{align}
 \d_\L \cY^\mu = i[\L,\bar\cY^\mu] = i[\xi_\nu(x) P^{\nu},P^\mu ] \sim 
g^{\mu\r}\del_\r \xi_\nu(x) P^\nu +  \xi_\nu(x) \cM^{\mu\nu} \ .
\end{align}
The last term can be dropped at $p$ (and it vanishes upon averaging $[ 
\cM^{\mu\nu}]_0 =0$ over $\cK$).
Then  for the symmetric part of $h_{\mu\nu}$,
the transformation law of a  graviton under diffeomorphisms is recovered.
\paragraph{String states.}
The above metric applies to semi-classical modes with small momenta.
Now recall the triple self-intersecting global structure as discussed in 
Section \ref{sec:global}.
This leads to an extra sector in the space $\End(\cH)$ of (non-commutative) 
functions,
which do not have any classical analog. They are best described as string states
$|p\rangle\langle p'| \in \End(\cH)$ where  $|p\rangle$, $|p'\rangle$ denote 
coherent states on  different sheets \cite{Steinacker:2016nsc}. 
Their energy from the Laplacian $\Box_\cY$ is proportional to the distance between $p$ and $p'$ in target space.
Hence for the above phase space embeddings, the states connecting  different sheets of $t$ space  
at the intersections have lowest energy, and
should lead to 3 generations as discussed in \cite{Steinacker:2015mia}. 
\subsection{Position space embeddings}
Analogously to the discussion of the momentum space embedding, we can 
investigate the effective metric for the two spherical embeddings 
found in Section \ref{sec:spherical-solutions}.

To start with, consider $ \cY^A = ( X^a , 0 )^T$. To emphasize the $S^4_N\times 
S^4_n \times S^2$ geometry, we switch to the coordinates $x_N$, $x_n$, and $t$. 
We obtain the effective metric $G_X^{\mu\nu}$ (in the adapted notation 
\eqref{eq:notation_metric}) to be
\begin{subequations}
\label{eq:XBack_all}
\begin{align}
G^{bc}(x_N,x_N) &= \frac{N^2}{4} g^{bc} -x_N^b x_N^c 
 \label{eq:XBack_XNXN}\; ,\\
G^{bc}(x_n,x_n)&= \frac{n^2}{4} g^{bc} -x_n^b x_n^c
\label{eq:XBack_XnXn} \; ,\\
G^{bc}(x_N,x_n)&= -\frac{N^2+n^2}{8}g^{bc} 
+\frac{1}{2} t^{bc} -x_N^b x_n^c
\; ,\\
G^{bc}(t,t)&= \left(\frac{N^2+n^2}{2} -2\cR^2 
\right)t^{bc} + \left(\cR^4 - \frac{(N^2-n^2)^2}{16}  \right) g^{bc} -t^b t^c
\; ,\\
G^{bc}(x_N,t)&= \frac{1}{4}(N^2+n^2-4\cR^2) m_N^{bc} 
+ \frac{N^2}{2} m_n^{bc} -x_N^b t^c
\; ,\\
G^{bc}(x_n,t) &= \frac{n^2}{2}  m_N^{bc} 
+ \frac{1}{4}(N^2+n^2-4\cR^2)  m_n^{bc} 
-x_n^b t^c \; .
\end{align}
\end{subequations}
To make the comparison to a sphere manifest, consider the $4$-sphere of radius 
$R_N=\frac{N}{2}$ embedded in $\R^5$ via $x^5 = \pm \sqrt{R_N^2 - \sum_{a=1}^4 
x^a x^a}$ for $x^a$, $a=1,\ldots,5$. The induced metric $g_{bc}$ and 
its inverse $g^{bc}$ (on the ``north'' hemisphere) read
\begin{align}
 g_{bc} = \delta_{bc} + \frac{x_b x_c}{R_N^2 - \sum_{a=1}^4 x^a x^a} \; 
,\qquad 
 g^{bc} = \delta^{bc} - \frac{x^b x^c}{R_N^2} \; .
\end{align}
Therefore, the effective metric component \eqref{eq:XBack_XNXN} is up to a 
conformal rescaling by $R_N^2$ the  metric of the sphere $S_N^4$ spanned by 
$x_N^a$; while the component \eqref{eq:XBack_XnXn} corresponds to the small 
sphere $S^4_n$ of radius $R_n=\frac{n}{2}$ spanned by $x_n^a$.

However, the mixed components of \eqref{eq:XBack_all} show an intricate 
geometry. In order 
to disentangle the structure it is instructive to consider the expressions in 
the limit \eqref{eq:limit_metric}, for which one readily obtains
\begin{subequations}
\label{eq:XBack_leading}
\begin{align}
G^{bc}(x_N,x_N) &= \frac{N^2}{4} g^{bc} -x_N^b x_N^c \sim O(N^2)
 \; ,\\
G^{bc}(x_n,x_n)&\sim O(1)\; ,\\
G^{bc}(x_N,x_n) &\sim O(N)\; ,
\\
G^{bc}(t,t) &=\left( \frac{N^2n^2}{4} + 4 \delta^2 + O(N)\right) g^{bc} 
-t^b t^c
\sim O(N^2) \; ,
\\
G^{bc}(x_N,t)&= -2\delta\ m_N^{bc} + \frac{N^2}{2} m_n^{bc} -x_N^b t^c
\sim O(N^2)\; ,
\\
G^{bc}(x_n,t) &\sim O(N) \; .
\end{align}
\end{subequations}
Clearly all mixed components vanish upon averaging
(i.e.\ for the lowest harmonics on  $\cK$). Furthermore, components involving 
$x_n$ are sub-leading in $N$, which is expected in the considered limit 
\eqref{eq:limit_metric}.

Next, we focus on the choice $ \cY^A = ( Y^a , 0 )^T$. To begin with we 
compute the effective metric $(G_\cY^{\mu\nu} )$ on the $4$-sphere, which reads 
in the notation \eqref{eq:notation_metric} as follows: 
\begin{subequations}
\label{eq:YBack_all}
\begin{align}
 G^{bc}(x,x) &\equiv \{y^a,x^b\}\{y_a,x^b\} = -t^{bc}-y^b y^c 
+\frac{N^2+n^2}{2} g^{bc}   \label{eff-metric-Y-embed-full} \\
&\stackrel{\eqref{eq:limit_metric}}{= } \left(R_N^2+O(n^2)\right)  g^{bc} 
- x^b x^c \; .
\end{align}
Again, in the limit $N\gg n\geq1$ the effective metric 
agrees with the conformally rescaled  metric of a $4$-sphere of radius 
$R_N$ embedded in $\R^5$. This is not surprising, as $x$ and $y$ become 
identical to $x_N$ in the considered limit \eqref{eq:limit_metric}. The 
remaining components of $(G_\cY^{\mu\nu} )$ read in this limit as follows:
\begin{align}
G^{bc}(t,t)&=
n^2 x^b x^c
\sim O(N^2) \; ,
\\
G^{bc}(y,y)&=
\left(\frac{N^2}{4} +O(N)\right) g^{bc}
-\left(1 + O\left(N^{-1}\right)\right) x^b x^c 
-\left( \frac{4}{N^2} +O\left(N^{-3}\right)  \right)t^b t^c 
\\
&\approx \left(\frac{N^2}{4} +O(N)\right) g^{bc}
-\left(1 + O\left(N^{-1}\right)\right) x^b x^c 
 \sim O(N^2)  \; , \nn \\
G^{bc}(x,t)& =
-t^b x^c
\sim O(N^2) \; ,
\\
G^{bc}(x,y)& =
\left(\frac{N^2}{4} +O(N)\right) g^{bc}
-\left(1 + O\left(N^{-1}\right)\right) x^b x^c 
-\frac{4}{N^2-n^2} t^bt^c \\
&\approx \left(\frac{N^2}{4} +O(N)\right) g^{bc}
-\left(1 + O\left(N^{-1}\right)\right) x^b x^c
\sim O(N^2) \nn \; , \\
G^{bc}(t,y)&=
-\left(1 + O\left(N^{-1}\right)\right)
x^b t^c 
\sim O(N^2) \; .
\end{align}
\end{subequations}
As in the $(X^a,0)$ embedding \eqref{eq:XBack_leading} and
in contrast to the $(T^a,0)$ embedding \eqref{eq:TBack_all}, 
the leading metric components \eqref{eq:YBack_all} are  of order in $N^2$ here.
However in $t$ space  the metric is degenerate, as anticipated 
in section \ref{sec:funct-on-S4}.

Finally, the effective metric for the combined embeddings of the form $ \cY^A = ( T^a , Y^a )^T$
is obtained simply as a sum of the metrics $G_T^{\mu\nu}$ 
\eqref{eff-metric-T-embed-full} and $G_Y^{\mu\nu}$ 
\eqref{eff-metric-Y-embed-full}.
Since this depends on the relative scales of the two contributions, we refrain from writing this down explicitly here.
However, it is very encouraging to observe that 
the contribution from $G_T^{\mu\nu}$ gives a large contribution $O(N^4)$ to the 
$t$ space \eqref{G_T(t,t)}, 
in contrast to the $O(N^2)$ contribution from both embeddings to the $x$ space. 
This suggests that a large separation of scales seems to arise naturally between the base $S^4$ and the 
fuzzy extra dimensions in the  $(Y,T)$ embedding, leading to a  
suppression of the corresponding harmonics in $t$ space.
This would provide the desired justification for dimensional reduction.

%
%
%
\section{Remarks and Conclusion}

In the course of the paper 
we have developed the description of a novel class 
of fuzzy spaces: the generalized fuzzy 4-sphere $\cS^4_\L$. In Section 
\ref{sec:twisted_bdl_coadj_orbit} we described the classical geometry of the 
underlying $SO(6)$ or $SU(4)$ coadjoint orbit $\cO_\L$. Based on the 
understanding of the 
basic fuzzy sphere $\cS_N^4$, we showed that the classical geometry of $\cO_\L$ 
is locally a twisted (or 
equivariant) bundle \eqref{S4-S4-bundle-proj} over some (basic) 4-sphere. 
Moreover, we found suitable sets of 
$SO(5)$-covariant coordinates $(x_N,x_n,t)$ or $(x,y,t)$ for the non-extremal 
case 
$t\cdot t\neq0$. While the $x_{N,n}$ provide an intuitive picture of a large and 
small 4-sphere, see for instance \eqref{xNn-matrix-class}; $x$ and 
$y$ are preferred coordinates, which arise as solutions for fuzzy dynamics 
governed by the corresponding Laplacian.
The extremal case $t\cdot t =0$ is best understood by employing the 
$SU(3)$-picture inherited from the basic fuzzy sphere. Similarly to earlier 
research \cite{Steinacker:2015mia,Steinacker:2014lma}, the geometry near $t=0$ 
is a triple self-intersection.

The fuzzy aspects of the generalized $4$-sphere have been investigated in 
Section \ref{sec:fuzzy_geometry}. We identified a suitable set of 
operators $\cM,\tilde{\cM},\tilde{\cT}$ that generate the algebra of 
functions $\End(\cH_\L)$ and computed their multiplication algebra 
\eqref{MNnO-algebra}. Again, 
two suitable descriptions for $SO(5)$ vector operators arise: either 
$(X_N,X_n,T)$ or $(X,Y,T)$, for which we determined the commutator relations  
\eqref{eq:commutators_XNXn}. The consistency of all fuzzy results with the
semi-classical relations provides a non-trivial check of all calculations.

The aforementioned descriptions allowed us to study embeddings into matrix 
models in Section \ref{sec:M-M-embeddings}. Besides the obvious 
solution given by the embedding of the Lie algebra generators $X^a$, we have 
found two additional solutions for Yang-Mills-type matrix models. As 
demonstrated in the spherical 
embedding \eqref{eqn:sol_spherical_XN+Xn}-\eqref{eqn:sol_spherical_XN-Xn}, the 
vector operators $X$, $Y$ are, in fact, eigenvectors of the corresponding matrix 
Laplacian (with possible $\cR^2$ anti-commutator terms). Another interesting 
scenario is the phase-space embedding, which includes the $T$ vector operator, 
which itself is an eigenvector \eqref{eq:phase_space_T} of the Laplacian. As 
mentioned earlier, $(X,Y,T)$ turn out to be preferred operators for matrix 
models, while $(X_N,X_n,T)$ have more accessible algebraic properties. 

In particular, we found a new solution \eqref{T-Y-embedding} of type $(Y,T)$ 
involving 10 matrices,
which can naturally be interpreted as 4-sphere
with self-intersecting fuzzy extra dimensions. This is probably the most sophisticated 
candidate available for obtaining 4-dimensional physics from the (Euclidean) 
IKKT model.
We worked out the effective metric on the full bundle $\cO_\L$ underlying $\cS^4_\L$ in Section \ref{sec:eff_metric_4d_physics}, 
which should eventually allow a justification for dimensional reduction and 
provide an understanding of possible mass gaps.
This was one of the open issues in \cite{Steinacker:2016vgf}.
Indeed, we find that a large separation of scales seems to arise naturally 
between the base $S^4$ and the 
fuzzy extra dimensions in the  $(Y,T)$ embedding.
However, we cannot make any statements about the scale parameters at this point. 
Determining these
would presumably require a 1-loop computation; hence,   
we only briefly glanced on the physical implications. 

One notable omission of this paper is the oscillator (or Jordan-Schwinger) 
construction of $\cS^4_\L$.
This would generalize the spinor realization of fuzzy $\cS^4_N$ in
\cite{Grosse:1996mz},
involving two instead of one set of spinorial creation- and annihilation 
operators. 
This construction is particularly useful for non-compact and Lorentzian analogs of $S^4$ 
\cite{Hasebe:2012mz,Grosse:2010tm}. 
However, to keep the paper within bounds we decided to postpone this 
construction to another paper.

The results and insights of this paper offer intersecting perspectives and provide a solid basis  
for future work. 
One important step would be a 1-loop analysis on $\cS^4_\L$ in the IKKT model,  to 
determine the dynamical scale parameters. 
This should be possible using the techniques in \cite{Steinacker:2016nsc}. 
Given these parameters, the low-energy physics on such a background could be studied,
starting with a refined fluctuation analysis along the lines of \cite{Steinacker:2016vgf}. 
Due to the extra structure provided by the self-intersecting extra dimensions,
this is expected to lead to a non-trivial and  physically interesting higher-spin theory
in 4 dimensions.

\paragraph{Acknowledgments.}

Useful discussions with
J. Barrett, M. Buric,
S. Fredenhagen, H. Kawai, K. Krasnov,
K. Mkrtchyan,   J. Nishimura,  P. Presnajder, S. Ramgoolam, and G. Zoupanos 
are gratefully acknowledged.
This work was supported by the Austrian Science Fund (FWF) grant
 P28590, and by the Action MP1405 QSPACE from the European Cooperation in Science and Technology (COST).

\appendix
%
%
\section{Fuzzy algebra}
\label{app:fuzzy_algebra}
\subsection{Characteristic equation}
\label{app:chareq}
Consider the $6\times 6$ matrix of operators
\begin{align}
  \cM &= (\cM)^{ab} = \sum_{c<d}\cM^{cd} (\l_{cd})^{ab} \equiv \cM^{ab} \qquad 
\in Mat(6,\cA)
\end{align}
Note that $(\l_{cd})^{ab} = \d^a_c\d^b_d - \d^a_d\d^b_c$ is anti-hermitian 
while 
the $\cM^{ab}$are hermitian generators.
We start from the identity
\begin{align}
 \cM  &= -\frac i2 \sum_{a<b} ((\cM_{ab} + i\l_{ab})^2 - \cM_{ab} \cM^{ab} + 
\l_{ab} \l^{ab}) \ .
 \label{cM-matrix-tensor-6}
\end{align}
To compute its characteristic equation for $\L = (N,0,n)$, we note that the 
6-dimensional representation 
has highest weight  $\L_2 =(0,1,0)$.
Then 
\begin{align}
V_{\L}\tens V_{\L_2} &= \oplus_i V_{\L + \nu_i} , \qquad \nu_i \in 
\{(0,1,0),(-1,0,1),(1,0,-1),(-1,1,-1)\} \ .
\end{align}
We consider $\Lambda= (N,0,n)$. The $6$-dimensional representation 
has highest weight $\Lambda_1=(0,1,0)$. Then
\begin{subequations}
\begin{align}
 V_{\Lambda} \otimes V_{\Lambda_1} &= \bigoplus_{j=1}^4 
V_{\Lambda + \lambda_j} \\
\lambda_1 &=(1,0,-1)  \; , \quad
\lambda_2 =(-1,0,1) \; , \quad 
\lambda_3 =(-1,1,-1) \; , \quad 
\lambda_4 =(0,1,0) \; .
\end{align}
\end{subequations}
Using the Killing metric 
\begin{align}
 (\L_i,\L_j) = \frac 14\begin{pmatrix}
                        3&2&1\\
                        2&4&2\\
                        1&2&3
                       \end{pmatrix}
\end{align}
we obtain
\begin{subequations}
\begin{align}
i\cM\lvert_{\cH_{\Lambda+\lambda_j}} &= 
(\lambda_j, \Lambda +\rho) - (\Lambda_1,\rho) \; ,\\
i\cM\lvert_{\cH_{\Lambda+\lambda_1}} &= -\frac{1}{2} (N-n)-2\; , \\
i\cM\lvert_{\cH_{\Lambda+\lambda_2}} &= \frac{1}{2} (N-n)-2 
 \; , \\
i\cM\lvert_{\cH_{\Lambda+\lambda_3}} &= -\frac{1}{2} (N+n)-3 \; , \\
i\cM\lvert_{\cH_{\Lambda+\lambda_4}} &= \frac{1}{2} (N+n) \; .
\end{align}
\end{subequations}
As consistency check we verify
\begin{equation}
 \sum_{j=1}^4 \cM\lvert_{\cH_{\Lambda+\lambda_j}} \cdot 
\mathrm{dim}(V_{\Lambda + \lambda_j})=0 \; .
\end{equation}
This gives the characteristic equation \eqref{char-eq-fuzzy}
\begin{align}
 \left(\left(i\cM+2\right)^2 - \frac{(N-n)^2}{4}\right)\left(\left(i\cM+\frac 
32\right)^2 - \frac{(N+n+3)^2}{4}\right) = 0 \ .
\end{align}
In contrast for $\L = (N,0,0)$, we would obtain 
\begin{align}
V_{\L}\tens V_{\L_2} &= \oplus_i V_{\L + \nu_i} , \qquad \nu_i \in 
\{(0,1,0),(-1,0,1)\} \; ,
\end{align}
which leads to the quadratic characteristic equation for fuzzy $\C P^3_N$ 
\cite{Grosse:2004wm}.
%
%
\subsection{Structure constants}
\label{sec:stucture_constants}
The structure constants of the multiplication algebra \eqref{MNnO-algebra} are 
as follows:
\begin{subequations}
\label{Rxyz-full}
\begin{align}
 R_{NN}^N &= 
 \frac{ (N-n)^2 - 14\det{\tilde{A}} }{ 8 \det{\tilde{A}} } 
 \; ,\\
R_{NN}^n &=
\frac{  2   \det{\tilde{A}} - (N-n)^2 }{8 \det{\tilde{A}} }
 \; ,\\
 R_{NN}^T &= 0 = R_{nn}^T  
 \; ,\\
R_{NN}^g &= \frac{1}{8}  \left( 2  \det{\tilde{A}}  -4c  \right) 
\; ,\\
R_{nn}^N &= 
\frac{ 2 \det{\tilde{A}} +  (N-n)^2 }{8 \det{\tilde{A}} }
 \; ,\\
R_{nn}^n &=
- \frac{  (N-n)^2 + 14  \det{\tilde{A}}  }{8 \det{\tilde{A}} }
\; ,\\
R_{nn}^g &=  
-\frac{1}{8} \left(2 \det{\tilde{A}}  +4  c \right)
\; ,\\
R_{Nn}^N &= 
 - \frac{ 2 \det{\tilde{A}} +  (N-n)^2  }{8 \det{\tilde{A}} }
\; ,\\
R_{Nn}^n &= 
-\frac{  2  \det{\tilde{A}}- (N-n)^2 }{8 \det{\tilde{A}} }
\; ,\\
R_{Nn}^T &= 
 - \frac{1}{2}
\; ,\\
R_{Nn}^g &= 0 
\; ,\\
R_{TN}^N &= 0 = R_{Tn}^n
\; ,\\
R_{TN}^n &=
\frac{1}{2} \left(2 c -  \det{\tilde{A}} \right)
\; ,\\
R_{TN}^T &= 
 - \frac{3}{2}  = R_{Tn}^T
\; ,\\
R_{TN}^g &=\frac{1}{2} \left(C_2[\Lambda]-3c\right) = R_{Tn}^g 
\; ,\\
R_{Tn}^N &= 
\frac{1 }{2}  \left(2c + \det{\tilde{A}}  \right)
\; ,\\
R_{TT}^N &= 
\frac{1}{4} \left( 3N^2 + 6N + 3n^2 + 6n - 2Nn +  6\det{\tilde{A}} \right) 
\; ,\\
R_{TT}^n &= 
\frac{1}{4} \left( 3N^2 +6N+3n^2+6n-2Nn - 6\det{\tilde{A}} \right) 
\; ,\\
R_{TT}^T &=
3
\; ,\\
R_{TT}^g &= 
\frac{1}{16} 
\left(4 n^2 N^2
+12 n N^2
+12 n^2 N
+N^2
+n^2
+26 n N
-12 N
-12 n
\right)
\; .
 \end{align}
\end{subequations}

%
\subsection{Laplacians}
\label{sec:laplacians}
The matrix Laplacian $\Box_\cY$ for an embedding $\cY^A$ has been defined in 
\eqref{eq:def_Laplacian}. In this appendix we provide the details for 
embeddings based on linear combinations of $X_N$, $X_n$, and $\cT$.
\paragraph{\texorpdfstring{$X_N$}{XN}-Laplacian.}
\begin{align}
 \Box_{X_N} X_N^b =&
 \frac{1}{8} \Bigg(
3 
+\frac{N^2 +9 N +n^2 +9 n +4 n N -28 c}{\det{\tilde{A}}} \\
&+\frac{25 N^2 +138  N+25 n^2 +138 n +42 Nn +168}{(N+n+4)(N+n+2)}
\Bigg)
X_N^b  \nn \\
&-\frac{ (N-n)^2 -6  \det{\tilde{A}}}{8  \det{\tilde{A}}^2}
\left( X_N^b \cR^2 + \cR^2 X_N^b \right) \nn \\
&-\frac{1 }{4} 
\bigg(
2   
+  \frac{ 8c}{(N+n+4)(N+n+2) }
-\frac{ 3 N^2+ 3 n^2 +6 N +6 n -2 nN }{ \det{\tilde{A}} }
\bigg)
X_n^b \nn \\
&+\frac{(N-n)^2 -6  \det{\tilde{A}} }{8 \det{\tilde{A}}^2 }
\left( X_n^b \cR^2 + \cR^2 X_n^b \right) 
\; ,\nn \\
 \Box_{X_N} X_n^b =&
\frac{1 }{8   }
 \bigg(
 9
 +\frac{16 c}{ \det{\tilde{A}}}  
 -   \frac{9 N^2 + 42 N + 9 n^2 +42 n+ 10Nn +40}{(N+n+4)(N+n+2)}
\bigg)
 X_N^b  \\
&+\frac{(N-n)^2 -6  \det{\tilde{A}} }{8  \det{\tilde{A}}^2 }
\left( X_N^b \cR^2 + \cR^2 X_N^b \right) \nn \\
&+c  \frac{  (N-n)^2 -4  \det{\tilde{A}} }{2 \det{\tilde{A}}^2  } 
X_n^b \nn \\
&- \frac{ (N-n)^2 - 6  \det{\tilde{A}} }{8  \det{\tilde{A}}^2 }
\left( X_n^b \cR^2 + \cR^2 X_n^b \right) 
\; ,\nn \\
 \Box_{X_N} T^b =& \frac{5 }{2}   T^b \; .
\end{align}
\paragraph{\texorpdfstring{$X_n$}{Xn}-Laplacian.}
\begin{align}
 \Box_{X_n} X_N^b =&
c \frac{ (N-n)^2 + 4 \det{\tilde{A}}}{2 \det{\tilde{A}}^2}
 X_N^b  \\
&-\frac{(N-n)^2 +6  \det{\tilde{A}} }{8  \det{\tilde{A}}^2 }
\left(X_N^b \cR^2 + \cR^2 X_N^b \right) \nn \\
& +\frac{1}{8}
\Bigg(
 9  
-\frac{16  c}{\det{\tilde{A}}} 
- \frac{ 9 n^2+10 n N+42 n+9 N^2+42 N+40 }{(N+n+4)(N+n+2)}  
\Bigg)
X_n^b \nn \\
&+\frac{(N-n)^2 + 6 \det{\tilde{A}} }{8 \det{\tilde{A}}^2 }
\left(X_n^b \cR^2 + \cR^2 X_n^b \right) 
\; ,\nn \\
 \Box_{X_n} X_n^b =&
 -\frac{1 }{4  } 
 \left( 2  + \frac{8 c }{(N+n+4)(N+n+2) }
 + \frac{3 N^2 +3 n^2 +6 N  +6 n -2N n }{ \det{\tilde{A}} }
\right) X_N^b \nn \\
& +\frac{(N-n)^2 +6  \det{\tilde{A}} }{8 \det{\tilde{A}}^2  }
\left(X_N^b \cR^2 + \cR^2 X_N^b \right)  \\ 
&+\frac{1 }{8  \det{\tilde{A}}^2}
\Bigg(
3  
 - \frac{N^2 +9 N +n^2 +9 n +4 n N - 28 c}{ \det{\tilde{A}}}
 \nn \\
&+\frac{ 25 N^2 +138 N + 25 n^2 +138 n +42 Nn +168 }{ (N+n+4)(N+n+2)} 
\Bigg)
X_n^b \nn \\
& -\frac{(N-n)^2 + 6 \det{\tilde{A}} }{8 \det{\tilde{A}}^2 }
\left(X_n^b \cR^2 + \cR^2 X_n^b \right) 
\; , \nn \\
 \Box_{X_n} T^b =& \frac{5}{2}  T^b  \; .
\end{align}
\paragraph{\texorpdfstring{$\cT$}{T}-Laplacian.}
\begin{align}
 \Box_{\cT} X_N^b =&
 \frac{3}{2} \left(  4c + \det{\tilde{A}} \right) X_N^b 
-\frac{3}{2} \left( X_N^b \cR^2 + \cR^2 X_N^b \right)  \\
& -\frac{1}{2 } \left(3N +3n + 2 Nn + 3 \det{\tilde{A}} \right)
X_n^b 
+\frac{3}{2 }
\left( X_n^b \cR^2 + \cR^2 X_n^b \right) 
\; ,\nn \\
 \Box_{\cT} X_n^b =&
 -\frac{ 1}{2 }  \left( 3 N +3 n +2 n N -3  \det{\tilde{A}}  \right) X_N^b 
 +\frac{3 }{2} 
\left( X_N^b \cR^2 + \cR^2 X_N^b \right)  \\ 
&+\frac{3}{2} 
\left( 4c - \det{\tilde{A}} \right) 
X_n^b
 -\frac{3}{2}
\left( X_n^b \cR^2 + \cR^2 X_n^b \right) 
\; ,\nn \\
 \Box_{T} T^b =&
\frac{1}{2}  \left(9N^2 +30N +9 n^2+30n+2Nn \right) T^b 
- 6 \left(  T^b \cR^2 + \cR^2  T^b  \right)  \; .
\end{align}
\paragraph{Mixed Laplacian.}
For the mixed Laplacian \eqref{mixed-laplace}, we obtain
\begin{align}
\Box_{mix} X_N^b  =&
\frac{1}{8 } \Bigg(
13 
- \frac{5 N^2+21 N + 5 n^2+21 n +4N n - 28c}{\det{\tilde{A}} }
  \\
&-\frac{13 N^2+54 N + 13 n^2 +54 n +10 Nn  +40 }{(N+n+4)(N+n+2)} 
\Bigg)
X_N^b  \nn \\
&+\frac{1}{(N+n+4)(N+n+2)}
\left( X_N^b \cR^2 +\cR^2 X_N^b \right) \nn \\
&+ \frac{1 }{8}
\Bigg(
7 
- \frac{9 N^2+21 N+ 9 n^2+21 n-4 Nn -28c}{\det{\tilde{A}}}
 \nn \\
& +\frac{N^2-18 N+ n^2 -18 n -14 N n -56}{(N+n+4)(N+n+2)}
\Bigg)
X_n^b \nn \\
&-\frac{1}{(N+n+4)(N+n+2)}
\left( X_n^b \cR^2 +\cR^2 X_n^b \right) 
\; , \nn \\
\Box_{mix} X_n^b=&
\frac{1}{8 } \Bigg(
 7 
+ \frac{9 N^2+21 N +9 n^2 +21 n -4 N n -28c}{\det{\tilde{A}}}   \\
&+\frac{N^2-18 N +n^2-18 n -14 n N -56}{(N+n+4)(N+n+2)} 
\Bigg)
X_N^b  \nn\\
&-\frac{1}{(N+n+4)(N+n+2)} 
\left( X_N^b \cR^2 +\cR^2 X_N^b \right) \nn \\
& + \frac{1}{8 }
\Bigg(
13 
+\frac{5 N^2+21 N +5 n^2 +21 n +4N n -28c}{ \det{\tilde{A}}}
 \nn \\
&- \frac{13 N^2 +54 N +13 n^2 +54 n +10 N n +40}{(N+n+4)(N+n+2)}
\Bigg)
X_n^b \nn \\
&+\frac{1}{(N+n+4) (N+n+2)} 
\left( X_n^b \cR^2 +\cR^2 X_n^b \right)  
\; ,\nn \\
\Box_{mix}  T^b=&   3  T^b \; .
\end{align}
%

\bibliographystyle{JHEP}
\bibliography{papers}

\end{document}